\documentclass[aps,prl,showpacs,reprint,floatfix]{revtex4-2}

\usepackage{amsmath,amssymb,mathtools, bbm, xcolor, hyperref}

\newcommand{\lambdamax}{\lambda_\mathrm{max}}
\newcommand{\lambdamin}{\lambda_\mathrm{min}}

\newcommand{\bham}{\beta \mathcal{H}}
\newcommand{\eps}{\varepsilon}

\newcommand{\bx}{\mathbf{x}}

\newcommand{\bk}{\mathbf{k}}
\newcommand{\ddx}{d^d \bx}
\newcommand{\br}{\mathbf{r}}
\newcommand{\ddkp}{\frac{d^d \bk'}{(2\pi)^d}}
\newcommand{\ddk}{\frac{d^d \bk}{(2\pi)^d}}

\begin{document}

\title{Critical Properties and Glass Transitions in Randomly Coupled Fields}

\author{Amer Al-Hiyasat}
\affiliation{Department of Physics, Massachusetts Institute of Technology, Cambridge, Massachusetts 02139, USA}
\begin{abstract}
    We study an assembly of $N$ scalar fields coupled via a quenched random interaction, equivalent to a spin glass whose spins are promoted to $d$-dimensional fields. When $N$ is large, the random coupling matrix produces a family of Gaussian universality classes whose critical exponents are determined by the behavior of the eigenvalue density near the spectral edge. For a Wigner matrix, the correlation length diverges at criticality but the susceptibility remains finite, and one-loop replica analysis supports an upper critical dimension of unity. In an exactly solvable spherical variant, the heat capacity jumps at the transition in $d>1$, and correlations remain pinned to their critical form throughout the glass phase, giving generic scale invariance despite the absence of a Goldstone mode.
    \end{abstract}
\maketitle

Biological systems often comprise a large number of distinct interacting components, with pairwise couplings that are too numerous to resolve individually. This high-dimensional complexity is naturally tackled by the theory of disordered systems, which has yielded novel insights into areas as diverse as community ecology~\cite{may1972will, bunin2017ecological}, reaction network dynamics~\cite{kauffman1986autocatalytic}, neuroscience~\cite{hansel1993solvable}, and complex mixtures~\cite{sear2003instabilities,jacobs2017phase, bunin2025evolutionary, shrinivas2021phase, parkavousi2025enhanced, dinelli2026random}, both in and out of equilibrium. Unlike the spins in a spin glass, however, the interacting components in biological systems are often themselves spatially extended fields, representing, for example, species abundance densities or chemical concentrations. This spatial structure has qualitative consequences: in disordered generalized Lotka--Volterra models, spatial fluctuations can profoundly reshape coexistence and abundance dynamics~\cite{pearce2020stabilization, bunin2021directionality, garcia2024interactions, de2025self, al2026spatiotemporal}; in biomolecular mixtures, interactions among many concentration fields can drive phase separation into compositionally distinct condensates~\cite{sear2003instabilities,jacobs2017phase, shrinivas2021phase}; and in active bacterial mixtures, random motility regulation can generate spatially segregated communities~\cite{dinelli2026random}.

Across this breadth of applications, a common mathematical structure recurs: a large number of spatially extended fields interact locally through a quenched random coupling that is uniform in space. Despite growing interest in such systems, theoretical progress has been limited by a lack of simple, tractable models. Little is known about their critical behavior, even in thermal equilibrium, leaving open the question of whether random coupling generates novel universality classes. In this Letter, we answer this question in the affirmative: We introduce a minimal model of $N$ randomly coupled $d$-dimensional scalar fields, whose equilibrium statistical mechanics are governed by the effective Hamiltonian \looseness=-1
\begin{multline} \label{eq:fullHam} \bham = \int \ddx \bigg \{\sum_{i=1}^N \left[\frac{r}{2}\phi_i^2 + \frac{\kappa}{2} \left(\nabla \phi_i\right)^2\right] + \mathcal{U}(\{\phi_i\})\\
+ \frac{1}{2}\sum_{i\neq j} J_{ij} \phi_i \phi_j\bigg\},
\end{multline}
where $J$ is a random symmetric matrix which is uniform in space and $\mathcal{U}$ is a stabilizing nonlinearity. In $d=0$, Eq.~\eqref{eq:fullHam}
reduces to a soft fully-connected spin glass~\cite{sherrington1975solvable, mezard1988spin},
whereas for $N=1$, it is a scalar Landau--Ginzburg
theory in $d$ dimensions~\cite{kardar2007statistical}. Otherwise, Eq.~\eqref{eq:fullHam} can be viewed as a phenomenological field theory for a microscopic system in which an identical spin glass is copied onto every site of a $d$-dimensional
lattice, with corresponding spins coupled ferromagnetically between neighboring sites.

As \(r\) is lowered, the model undergoes a second-order transition at some \(r_c\), below which the uniform \(\phi_i=0\) state becomes unstable. In the following, we characterize the critical properties of this transition, working in the limit where the number of fields is taken to infinity \(N\rightarrow\infty\), before taking the system size large, \(L\rightarrow\infty\). We first show that the Gaussian critical behavior is controlled by the eigenvalue density of $J$ near its lower spectral edge, with free energy singularities that map onto those of an ordinary Gaussian theory in a higher effective dimension. For a Wigner matrix, the critical susceptibility remains finite, but the structure factor develops a \(|k|\) cusp. A replica calculation shows that one-loop corrections are infrared finite in \(d>1\), supporting an upper critical dimension \(d_u=1\). To characterize the transition from both sides, we then introduce an exactly solvable spherical variant. We find that its effective mass remains pinned at criticality throughout the glass phase, producing generic scale invariance without a Goldstone mode, and that space turns the continuous cusp in the heat capacity into a discontinuous jump. Together, our results introduce a novel class of spatially extended randomly-interacting models with broad potential applications, whose equilibrium statistical mechanics are both unconventional and analytically tractable. \looseness=-1

\paragraph{Gaussian theory.} Let us first consider the Gaussian theory obtained by setting $\mathcal U=0$, which is stable only in the high temperature phase, $r>r_c$. We may diagonalize $J$, denoting the ordered eigenvalues $\lambda_1 < \dots < \lambda_N$, and obtain a theory of $N$ independent eigenmodes, each with mass $(r + \lambda_i)$. The Gaussian critical point is located at $r_c = -\lambda_1$, where the first eigenmode becomes massless. To obtain a well-defined large-$N$ limit, we demand of $J$, without specifying its detailed distribution, that its eigenvalue density has a large $N$ limit, $\rho(\lambda) \equiv \lim_{N\rightarrow \infty} \frac{1}{N} \sum_{i=1}^N \delta(\lambda - \lambda_i)$, and that $\rho(\lambda)$ is compactly supported on a finite interval $(\lambda_{\mathrm{min}}, \lambda_{\mathrm{max}})$. The critical $r$ then concentrates at $r_c = - \lambdamin$ for large $N$. This invites the definition of a reduced temperature parameter setting the distance to criticality,
\[\tau \equiv r + \lambdamin.\]
The critical properties near $\tau=0$ will depend on the behavior of $\rho(\lambda)$ near its lower edge. We assume the general form~\cite{marchenko1967distribution, pastur1972spectrum}
\begin{equation} \label{eq:edge} \rho(\lambdamin + \eps) \sim A \varepsilon^\theta, \qquad \varepsilon \rightarrow 0^+,
\end{equation}
 with $A>0$ and $\theta > -1$ to ensure normalizability. The case $\theta = 1/2$ is of particular interest, as it includes the Gaussian Orthogonal Ensemble (GOE), for which $\rho(\lambda)$ is a Wigner semicircle~\cite{potters2020first}. Examples of both $\theta=1/2$ and $\theta = -1/2$ are provided by full-rank Wishart ensembles~\cite{marchenko1967distribution}.

The partition function, $Z = \int \prod_i \mathcal{D} \phi_i e^{-\bham}$, is readily evaluated as a Gaussian integral, from which we obtain the free energy density
\begin{equation}\beta f \equiv - \frac{\log Z}{N V} = \frac{1}{2} \int_{\lambdamin}^{\lambdamax} d\lambda \rho(\lambda) \int \ddk \log(r + \kappa k^2 + \lambda),\end{equation}
where the limit $N\rightarrow \infty$ has been taken to replace the sum over eigenvalues with a $\rho(\lambda)$-weighted integral. The singular contribution to the free energy, from which the critical properties originate, is dominated by the spectral edge behavior defined in Eq.~\eqref{eq:edge}; we write this in the following form
\begin{align} 
\beta f^{\mathrm{sing}} &\propto \int_0^{\Delta \lambda} d\eps \int \ddk \, \eps^{\theta} \log(\tau + \kappa k^2 + \epsilon), \nonumber \\
&\propto \int_0^{\Delta \lambda} dp \int_0^\Lambda dk \, p^{2\theta+1} k^{d-1} \log(\tau + \kappa k^2 + p^2) \label{eq:effectiveDimension}
\end{align}
where $\Delta\lambda \equiv \lambda_\mathrm{\max}-\lambdamin$, and $\Lambda$ is an ultraviolet (UV) cutoff. In the second line, we have made the change of variables $p \equiv \sqrt{\eps}$, which reveals that the singular part of the free energy is identical in form to that of a single Gaussian field in dimension $d_\mathrm{eff} \equiv d + 2 \theta + 2$, with anisotropy in the $2\theta + 2$ additional dimensions~\cite{van2010second}. This is suggestive of an upper critical dimension
\begin{equation} \label{eq:dupper}
    d_u = 2(1-\theta),
\end{equation}
obtained by setting $d_\mathrm{eff} = 4$. The effect of a ``soft" edge ($\theta>0$) is thus to tame the critical singularities, whereas a diverging ``hard" edge ($\theta<0$) enhances them. In the Wigner case $\theta = 1/2$, we predict $d_u = 1$, which will soon be verified through an explicit one-loop computation in the replica picture. This implies the Gaussian critical exponents obtained in this section to apply in all $d>1$ when $J$ is a GOE random matrix.

All $\tau$ derivatives of $\beta f^{\mathrm{sing}}$ will behave as they do in the corresponding $d_\mathrm{eff}$-dimensional Gaussian theory: The critical heat capacity, $C(\tau) \equiv -\partial_\tau^2 \beta f \propto \tau^{-\alpha}$, is finite above $d_u$ and divergent below it, so that
\begin{equation} \alpha =\begin{cases} 1-\theta - \frac{d}{2}, & d < d_u, \\
 0, & d>d_u.
\end{cases}
\end{equation} 

The remaining critical exponents do not follow trivially from those of the $d_\mathrm{eff}$-dimensional theory, as they are defined in terms of observables which probe only the original $d$-dimensional subspace. For example, the two-point correlation function takes the form
\begin{equation}\langle \phi_i(\bk) \phi_j(\bk') \rangle = (2\pi)^d 
\delta(\bk+\bk') \left[(r+\kappa k^2)\mathbbm{1}+J\right]_{ij}^{-1}.\end{equation}
The matrix on the right hand side is related to the resolvent of $J$, defined as $R(z) \equiv (z \mathbbm{1} - J)^{-1}$. For an orthogonally-invariant ensemble, $R$ self-averages to a diagonal form: $\lim_{N\rightarrow \infty} R_{ij}(z) = \overline{R_{ij}(z)} = \delta_{ij} g(z)$, where $g(z)$ is the Stieltjes transform of $\rho(\lambda)$ \cite{potters2020first}. Using these results, the structure factor, defined via $\langle \phi_i(\bk) \phi_j(\bk') \rangle \equiv (2\pi)^d 
\delta_{ij} \delta(\bk+\bk') S(k)$, may be expressed for large $N$ as
\begin{align}
    S(k) &=  \int_{\lambdamin}^{\lambdamax}  \frac{d\lambda \rho(\lambda)}{r + \kappa k^2 + \lambda} \simeq \int_0^{\Delta \lambda} \frac{d\eps A \eps^\theta}{\tau + \kappa k^2 + \eps} \nonumber \\
    &\simeq c - \frac{A \pi}{\sin(\pi \theta)}(\tau + \kappa k^2)^\theta + \mathcal{O}(\tau+\kappa k^2),  \label{eq:structureFactor}
\end{align}
where we have isolated the dominant contribution for small $(\tau + \kappa k^2)$, and $c$ is a nonsingular quantity that depends on the details of $\rho(\lambda)$.  Notably, for $\theta \geq 0$, the susceptibility, $\chi \equiv S(k=0)$, is finite at criticality, whereas for $\theta<0$ it diverges. Writing $\chi(\tau) \sim \tau^{-\gamma}$, we find
\begin{equation}
    \gamma = \begin{cases}
        |\theta|, &-1<\theta<0, \\
        0, &\theta>0.
    \end{cases}
\end{equation}
with logarithmic behavior when $\theta = 0$. Even where $\chi(0)$ is finite, the critical structure factor remains nonanalytic at the origin due to terms $\propto |k|^{2\theta}$. This induces a diverging correlation length and a critical power law: computing the inverse transform of Eq.~\eqref{eq:structureFactor}, we find~\cite{si}, up to subleading short-ranged corrections,
\begin{equation}\langle \phi_i(\bx) \phi_j(\bx') \rangle \sim \frac{\delta_{ij}}{|\bx - \bx'|^{d+2\theta}} f_\theta\left(\frac{|\bx - \bx'|}{\xi}\right), \,\,\quad \xi\equiv \sqrt{\frac{\kappa}{\tau}}. \label{eq:realspacecorr}\end{equation}
where $f_{\theta}(y)$ may be expressed in terms of Bessel functions and goes to a constant for $y \rightarrow 0$, while decaying exponentially as $f_\theta(y) \sim y^{(d+2\theta-1)/2} e^{-y}$ when $y\gg 1$. This form assumes $d>-2\theta$, else the inverse transform is infrared (IR)-divergent at criticality. Using the conventional definitions $\xi \sim \tau^{-\nu}$ and $\langle \phi_i(\bx) \phi_i(\bx')  \rangle_{\tau=0} \sim |\bx - \bx'|^{-(d-2+\eta)}$, we conclude
\begin{equation}
    \nu = \frac{1}{2}, \qquad \eta = 2 + 2 \theta.
\end{equation}
For a GOE (Wigner) matrix, defined via
\begin{equation} \label{eq:goe} \overline{J_{ij}} = 0, \qquad \overline{J_{ij} J_{k\ell}} = \left(\delta_{i k} \delta_{j \ell} +\delta_{i \ell} \delta_{j k} \right) \frac{\sigma^2}{N},\end{equation}
the eigenvalue density is supported on $(-2\sigma, 2\sigma)$ and follows the semicircular form, $\rho(\lambda) = \frac{1}{2\pi\sigma^2}\sqrt{4\sigma^2-\lambda^2}$. The Stieltjes transform of the latter is explicitly calculable, yielding
\begin{equation} \label{eq:wignersf} S(k) = \frac{2}{r+\kappa k^2+\sqrt{(r+\kappa k^2)^2-4\sigma^2}}.
\end{equation}
This behaves as $S(k) \simeq \sigma^{-1} - \sigma^{-3/2} \sqrt{\tau + \kappa k^2}$ near criticality, in accord with Eq.~\eqref{eq:structureFactor}, implying $\gamma = 0$ and a $|k|$ critical cusp. It follows that the critical correlation function decays in real space as a short-ranged power law $\propto x^{-d-1}$, leading to the unusually large exponent $\eta = 3$.

\paragraph{Replica analysis.} We next study the stability of Gaussian exponents to a quartic interaction term of the form
\begin{equation} \label{eq:phi4nonlinearity}
    \mathcal U(\{\phi_i\}) = \sum_{i=1}^N \frac{u}{4} \phi_i^4.
\end{equation}
For simplicity, we restrict to the GOE case defined in Eq.~\eqref{eq:goe}. Under these choices, Eq.~\eqref{eq:fullHam} defines a collection of randomly coupled scalar $\phi^4$ models. When $\sigma^2=0$, the independent fields have a critical point in the Ising universality class with $d_u = 4$. For $\sigma \neq 0$, the zero-dimensional model is a soft Sherrington Kirkpatrick (SK) spin glass~\cite{sompolinsky1982relaxational}. The full $d$-dimensional model shares symmetries with the models considered in Refs.~\cite{sear2003instabilities, jacobs2017phase, shrinivas2021phase} to study phase-separating multicomponent mixtures.

To study the impact of the nonlinearity, direct perturbation theory in the eigenbasis of $J$ is cumbersome, as $\mathcal U(\{\phi_i\})$ couples the eigenmodes through the random eigenvectors of $J$. It is more convenient to work in the replica formalism, in which the free energy is evaluated using the identity $\overline{\log Z} = \lim_{n\rightarrow 0} (\overline{Z^n}-1)/n$.  We show in the Supplemental Material~\cite{si} that the replicated partition function can be expressed as a functional integral over a bilocal matrix-valued field, $Q_{\alpha \beta} (\bx, \bx')$, in the form $\overline{Z^n} = \int \mathcal{D} Q \, e^{-N \mathcal{S}[Q]}$, where
\begin{align}\label{eq:action}
    \mathcal{S} \equiv \frac{\sigma^2}{4} &\sum_{\alpha, \beta = 1}^n \int_{\bx, \bx'} Q^2_{\alpha \beta}(\bx, \bx') - \log\left[\int {\prod_{\alpha=1}^n} \mathcal{D} \phi^\alpha  e^{-\bham_n}\right].
\end{align}
In the above, we have used the shorthand $\int_{\bx, \bx'} \equiv \int \ddx \ddx'$, and $\bham_n[\{\phi^\alpha\}; Q]$ is an effective Hamiltonian which couples the replica fields $\phi^1(\bx), \dots, \phi^n(\bx)$,
\begin{align} \label{eq:singlesitemeasure}
    \bham_n =& \sum_{\alpha=1}^n \int_\bx \left[ \frac{r}{2}(\phi^\alpha)^2 + \frac{\kappa}{2} \left(\nabla \phi^\alpha\right)^2 + \frac{u}{4} (\phi^\alpha)^4 \right]\nonumber \\
    &- \frac{\sigma^2}{2} \sum_{\alpha, \beta=1}^n \int_{\bx, \bx'} Q_{\alpha \beta} (\bx, \bx') \phi^\alpha(\bx) \phi^\beta(\bx').
\end{align}
For large $N$, $\overline{Z^n}$ may be evaluated by the saddle point method; extremizing $\mathcal{S}[Q]$ in Eq.~\eqref{eq:action} yields the self-consistent condition
\begin{equation} \label{eq:selfconsistency}
    Q_{\alpha \beta}(\bx, \bx') = \langle \phi^{\alpha}(\bx) \phi^{\beta}(\bx') \rangle_{\bham_n}
\end{equation}
where $\langle \cdot \rangle_{\bham_n}$ denotes an average with respect to the Boltzmann measure of the Hamiltonian in Eq.~\eqref{eq:singlesitemeasure}. For large $N$, Equation~\eqref{eq:selfconsistency} identifies $Q_{\alpha \beta}$ as the replica overlap matrix, as in a classical fully-connected spin glass. Here, however, $Q_{\alpha \beta}(\bx, \bx')$ is promoted to a bilocal function of space, since the replicas are themselves fields. Similar structure arises in the replica theory of quantum spin glasses, with time replacing the space variables~\cite{bray1980replica}. By translational invariance, we may write $Q_{\alpha \beta}(\bx, \bx')$ as a function of the separation, $Q_{\alpha \beta}(\bx - \bx')$, satisfying the symmetry $Q_{\alpha \beta}(\br) = Q_{\beta \alpha}(-\br)$.

Working first at the mean field level, we consider spatially uniform configurations of the form $Q_{\alpha \beta}(\br) \equiv q_{\alpha \beta}, \phi^\alpha(\bx) \equiv m^\alpha$~\cite{si}. The theory then has the structure of a soft-spin SK model, but with thermal fluctuations suppressed in the large volume limit $V \equiv \int d^d \bx \rightarrow \infty$: Above a critical $r_c = 2\sigma$, the replica-symmetric (RS) paramagnetic solution $m^\alpha = q_{\alpha \neq \beta} = 0$ is stable. For $r<r_c$, the RS spin glass solution is unstable~\cite{si, de1978stability}, and, by analogy to the SK model, the system is expected to exhibit full Parisi replica symmetry breaking (RSB)~\cite{mezard1988spin}. In the following, we consider the effect of spatial fluctuations about the uniform configuration, working perturbatively in powers of $u$.

\paragraph{Spatial fluctuations and perturbation theory.} For $r > r_c$, we make the RS diagonal ansatz 
\begin{equation} Q_{\alpha \alpha}(\bx - \bx') \equiv C(\bx - \bx'), \qquad Q_{\alpha\neq \beta} = 0,
\end{equation}
with $C(\bx - \bx')$ denoting the two-point correlation function of a single replica, according to Eq.~\eqref{eq:selfconsistency}. The fields then decouple in Eq.~\eqref{eq:singlesitemeasure}, so that a representative replica is governed by the self-consistent Hamiltonian
\begin{equation} \label{eq:highTemp}
    \bham_{\mathrm{eff}} = \frac{1}{2}\int \ddk \left[r+\kappa k^2 - \sigma^2 S(k)\right]|\phi(\bk)|^2 + U_{\mathrm{int}},
\end{equation}
where $S(k)$ is again the structure factor, obtained from the Fourier transform of $C(\bx - \bx')$, and $U_{\mathrm{int}} \equiv \frac{u}{4} \int \ddx \phi^4$. Unlike Eq.~\eqref{eq:fullHam}, Eq.~\eqref{eq:highTemp} is directly amenable to diagrammatic perturbation theory; the only complication over standard $\phi^4$ theory is that the Gaussian propagator must be determined self-consistently at each order.

At the bare level $(u=0)$, setting $\langle \phi(\bk) \phi(\bk') \rangle_0 = (2\pi)^d \delta(\bk + \bk') S_0(k)$ yields $S_0(k) = [r + \kappa k^2 - \sigma^2 S_0(k)]^{-1}$. This is Dyson's well-known equation for the resolvent of a Wigner matrix~\cite{pastur1972spectrum}, and its solution recovers Eq.~\eqref{eq:wignersf}. To one loop order, the self consistent equation for $S(k)$ takes the form
\begin{equation} \label{eq:highTempSkselfcon} S^{-1}(k) + \sigma^2 S(k) = r + \kappa k^2  + 3 u \int \ddkp S_0(k') + \mathcal{O}(u^2).\end{equation}
This amounts to a mass renormalization $r_\mathrm{eff}(r) = r + 3u \int \ddkp S_0(k')$, with the new critical point determined by $r_{\mathrm{eff}}(r_c) = 2\sigma$. $S(k)$ is then given by Eq.~\eqref{eq:wignersf}, but with $r_\mathrm{eff}$ in place of $r$, so that the critical susceptibility remains finite and unrenormalized at this order, $S(0)|_{r=r_c} = \sigma^{-1} + \mathcal{O}(u^2)$. The loop correction produces no leading critical singularities: the singular contribution to the derivative $\partial_r r_{\mathrm{eff}} |_{r=r_c}$ is proportional to the integral $\int_0^\Lambda dk k^{d-2}$, which is IR-finite in $d>1$ and logarithmically divergent in the marginal dimension $d=1$~\cite{si}. The renormalized mass $\tau_{\mathrm{eff}}(r) = r_{\mathrm{eff}}(r) - 2\sigma$ thus vanishes linearly with $(r-r_c)$ in $d>1$, so that the Gaussian  high-temperature exponents are unchanged. This supports the conclusion of Eq.~\eqref{eq:dupper} that the upper critical dimension is $d_u = 1$. We stress the distinction from finite-dimensional spin glasses, for which $d_u = 6$~\cite{harris1976critical}.

In the spin glass phase $r<r_c$, the replica-symmetric saddle becomes unstable~\cite{si, de1978stability}, and determining the ordered-side critical behavior requires characterizing the resulting full replica-symmetry-breaking problem, which lies beyond the scope of the present work. To obtain a controlled characterization of the glass transition from both sides, we instead introduce a spherical variant which is exactly solvable for large $N$.

\paragraph{Spherical model.} We now consider Eq.~\eqref{eq:fullHam} with an $O(N)$-symmetric nonlinearity of the form $\mathcal{U}(\{\phi_i\})= u\big[\sum_{i=1}^N \phi_i^2 \big]^2/4N$.
In vector form $\vec{\phi}\equiv (\phi_1, \dots, \phi_N)$, the Hamiltonian reads 
\begin{equation} \label{eq:sphericalHam} \bham =  \int \ddx \left[\frac{r}{2}\vec{\phi}^{\,2} + \frac{\kappa}{2} (\nabla \vec{\phi})^2 + \frac{1}{2}\vec{\phi}^{\,T} J \vec{\phi} + \frac{u}{4N} (\vec{\phi}^{\,2})^2\right]\end{equation}
The factor of $1/N$ in the nonlinearity keeps all terms in Eq.~\eqref{eq:sphericalHam} of order $N$. For $J=0$, this is the field theory of the $O(N)$ model, which for $N\rightarrow \infty$ describes the spherical ferromagnet~\cite{zinnjustin, berlin1952spherical}. Nonzero $J$ breaks $O(N)$ symmetry, so that Eq.~\eqref{eq:sphericalHam} may be regarded as a spherical $O(N)$ model with randomly anisotropic mass.

The zero dimensional limit of this model is much simpler than that of Eq.~\eqref{eq:phi4nonlinearity}; whereas the latter corresponds to a soft SK model, the present model maps to a soft spherical spin glass in the large-$N$ limit~\cite{kosterlitz1976spherical, nieuwenhuizen1985spherical}. As $r$ is lowered below $r_c$, the equilibrium measure condenses onto the lowest eigenmode of $J$, and the resulting glass phase is replica-symmetric. At the glass transition, the heat capacity is continuous but has a cusp, and the Edwards-Anderson overlap $q = N^{-1} \sum_{i=1}^N \langle \phi_i \rangle^2$ grows linearly in $(r_c - r)$~\cite{edwards1975theory}.

To study the role of spatial fluctuations, we solve Eq.~\eqref{eq:sphericalHam} exactly in the large-$N$ limit using the replica formalism. Introducing the overlap field $Q_{\alpha \beta}(\bx - \bx')$ decouples both the replica interactions and the $(\vec{\phi}^{\,2})^2$ term simultaneously, yielding the same saddle point condition as Eq.~\eqref{eq:selfconsistency} but with a Gaussian $\bham_n$~\cite{si}.
%\begin{align}
%\bham_n={}&\sum_\alpha \int_\bx\left[\frac{1}{2}\left[r-uQ_{\alpha\alpha}(0)\right](\phi^\alpha)^2+\frac{\kappa}{2}(\nabla\phi^\alpha)^2\right]\nonumber\\&+
%\frac{\sigma^2}{2}\sum_{\alpha\beta}\int_{\bx,\bx'}Q_{\alpha\beta}(\bx -\bx')\phi^\alpha(\bx)\phi^\beta(\bx').
%\end{align}
The correlator $\langle \phi^{\alpha}(\bx)\phi^{\beta}(\bx') \rangle_{\bham_n[Q]}$ can then be evaluated exactly, yielding an explicit self-consistent equation for $Q$. In Fourier space, this reads
\begin{equation}\label{eq:sphericalOverlapSelfcon}
    Q^{-1}_{\alpha \beta}(\bk) + \sigma^2 Q_{\alpha \beta}(\bk) = (r+\kappa k^2+ u p_\alpha) \delta_{\alpha \beta},
\end{equation}
where $p_\alpha(r) \equiv \int \ddk Q_{\alpha \alpha}(\bk)$ is a local diagonal overlap, whose real space expression is $p_\alpha = Q_{\alpha\alpha}(\bx - \bx' = 0) = \langle (\phi^\alpha)^2\rangle$. We solve Eq.~\eqref{eq:sphericalOverlapSelfcon} under a RS ansatz $Q_{\alpha \beta}(\bk) = S(k) \delta_{\alpha \beta} + q(k) (1-\delta_{\alpha \beta})$, which can be proven stable at all temperatures~\cite{si}. In the high temperature phase, where $q = 0$, the structure factor $S(k)$ takes the Gaussian form of Eq.~\eqref{eq:wignersf}, but with $r$ shifted self-consistently into $r_\mathrm{eff} = r + u p(r)$. As before, the shifted mass $\tau_{\mathrm{eff}} \equiv r + u p(r) - 2\sigma$ can be verified to vanish linearly with $(r-r_c)$ in $d>1$, so that the high-temperature critical exponents are identical to those of the Gaussian model. 

In the low temperature phase, it can be proven that the off-diagonal overlap is uniform in space $Q_{\alpha \neq \beta}(\bx - \bx') = q$, implying that different replicas couple only through their zero modes~\cite{si}. The overlap grows linearly below the critical point, as in the zero-dimensional model~\cite{kosterlitz1976spherical},
\begin{equation}
    q = \frac{r_c - r}{u}.
\end{equation}
This implies that spatial fluctuations do not renormalize the critical exponent characterizing the vanishing of the order parameter. In terms of the original (unreplicated) fields, $q$ is interpreted as an Edwards-Anderson order parameter $q = \overline{\langle \phi \rangle^2}$, suggesting that a typical $\langle \phi_i \rangle$ vanishes as $(r_c - r)^{1/2}$ near criticality. The diagonal local overlap $p$ instead gives $\overline{\langle \phi^2\rangle}$, and behaves as $p = (2\sigma - r)/u$ in the low temperature phase. Notably, this means that the mass, $\tau = r + up(r) - 2\sigma$, remains pinned at zero, so that connected correlations retain their critical form all throughout the glass phase \cite{si}. Indeed, for $r<r_c$, we find
\begin{equation}
    S(k) = \frac{2}{2\sigma + \kappa k^2 + \sqrt{(2\sigma + \kappa k^2)^2 - 4\sigma^2}} + q (2\pi)^d \delta^{(d)}(k).
\end{equation}
The first term contributes the $|k|$ critical cusp responsible for long-ranged correlations $\overline{\langle \phi(\bx) \phi(\bx') \rangle_c} \propto |\bx - \bx'|^{-d-1}$. Similar behavior is observed in the low temperature phase of the spherical ferromagnet~\cite{zinnjustin}. There, however, the long-ranged correlations result from the Goldstone mode associated with $O(N)$ symmetry breaking (i.e. the massless transverse fluctuations of the order parameter). Here, Eq.~\eqref{eq:sphericalHam} is not $O(N)$-symmetric for any particular realization of $J$, even though the GOE ensemble is itself $O(N)$-invariant. This is thus an example of an equilibrium model with short-ranged interactions that displays generic scale invariance without a Goldstone mode. 

Having computed the overlap field, we may now evaluate the quench-averaged free energy via the replica trick. Defining $\beta f \equiv -\overline{\log Z}/(NV)$, we find~\cite{si}:
\begin{equation}
    \beta f = \frac{1}{4}\int\ddk\left[\sigma^2S_{\mathrm{conn}}^2(k)-2\log S_{\mathrm{conn}}(k)\right]-\frac{u}{4}p^2.
\end{equation}
where $S_{\mathrm{conn}}(k) = S(k) - q (2\pi)^d \delta^{(d)}(k)$ is the Fourier transform of the connected correlation function. Differentiating twice with respect to $r$ yields the heat capacity $C(r)$; in both phases, this simplifies to $C(r) \equiv -\partial_r^2 \beta f = -\partial_r p(r)/2$.  Notably, $C(r)$ takes different limits on approaching criticality from either side
\begin{equation}
    C(r) = \begin{dcases}
        (2u)^{-1}, &r<r_c \\
        (2 u + 4\sigma^2/B)^{-1}, &r \rightarrow r_c^+
    \end{dcases}
\end{equation}
where $B$ is a momentum integral that is finite in $d>1$ and logarithmically divergent in $d=1$~\cite{si}. For any $d>1$, therefore, the heat capacity is discontinuous at the glass transition. This is different from the continuous cusp typically observed in fully-connected spin glasses, or in the $d=0$ limit of the present model~\cite{kosterlitz1976spherical}.

\paragraph{Discussion.} In this work, we studied the critical properties of a system with many randomly interacting, spatially extended components. Models of this form have arisen repeatedly in specialized settings~\cite{sear2003instabilities, jacobs2017phase, shrinivas2021phase, bunin2025evolutionary, dinelli2026random, hansel1993solvable, de2025self}, but have rarely been studied in general, in part due to a lack of minimal and analytically tractable models. The family of models introduced here provides such a framework.

In the Gaussian theory, we identified novel universality classes whose critical exponents are controlled by the eigenvalue density of the coupling matrix near its spectral edge. This behavior requires taking $N\rightarrow\infty$ before the thermodynamic $(L\rightarrow\infty)$ or critical $(r\rightarrow r_c)$ limits. At finite $N$, the spectrum is discrete and the critical point is sample-dependent, $r_c(N)=-\lambda_1$. Sufficiently close to $r_c(N)$ and at sufficiently long wavelengths, the lowest eigenmode is isolated and the behavior crosses over to that of an ordinary critical single field. The spectral-edge theory therefore applies outside a Ginzburg interval whose width vanishes with $N$. %describes a large-\(N\) regime that extends increasingly close to criticality and to longer length scales as \(N\) is increased~\cite{si}.

To move beyond the Gaussian level, we introduced two nonlinear versions: The component-wise $\phi^4$ model of Eq.~\eqref{eq:phi4nonlinearity} has the soft SK model as its zero-dimensional counterpart, whereas the spherical model of Eq.~\eqref{eq:sphericalHam} maps to the $p=2$ spherical spin glass for $N$ large and $d=0$~\cite{kosterlitz1976spherical}. In the former case, replica analysis allows the high temperature phase to be characterized perturbatively, but the low-temperature RSB phase remains to be solved. The spherical model, in contrast, is exactly solvable at all temperatures: its off-diagonal overlap remains uniform and grows linearly below criticality, connected correlations remain scale-free throughout the glass phase, and fluctuations cause the heat capacity to jump at the transition. It remains an open question to determine which of these features extend to the component-wise $\phi^4$ model, likely a more realistic representation of many-component mixtures~\cite{sear2003instabilities, jacobs2017phase, shrinivas2021phase}.

An interesting direction for future work is to study the dynamics of these models following a high temperature quench, where the interplay of glassy aging with critical slowing down and coarsening are likely to produce highly nontrivial dynamics. The spherical variant introduced here provides a promising tractable starting point for investigating both conserved and nonconserved dynamics, as well as the role of domain walls~\cite{cugliandolo1995full}.

\begin{acknowledgments}
\paragraph{Acknowledgments.} I am grateful to Mehran Kardar, Frédéric van Wijland, Rob Jack, and Guy Bunin for helpful discussions, and to Julien Tailleur for a critical reading of the manuscript. I thank the Kavli Institute for Theoretical Physics (KITP) for its hospitality. This work was supported in part by a travel grant from the
Institute for Complex Adaptive Matter (ICAM), and by grant NSF PHY-2309135 to the KITP.
\end{acknowledgments}
\bibliography{bibliography}
\end{document}

\title{Supplemental Material for ``Critical Properties and Glass Transitions in Randomly Coupled Fields"}

\author{Amer Al-Hiyasat}
\affiliation{Department of Physics, Massachusetts Institute of Technology, Cambridge, Massachusetts 02139, USA}

\maketitle 

\tableofcontents
\section{Gaussian theory} \label{sec:gaussian}
We consider here Eq.~\eqref{main-eq:fullHam} of the main text in the Gaussian case $\mathcal U = 0$,
\begin{equation} \label{eq:siGaussianHam} \bham = \int \ddx \bigg \{\sum_{i=1}^N \left[\frac{r}{2}\phi_i^2 + \frac{\kappa}{2} \left(\nabla \phi_i\right)^2\right] + \frac{1}{2}\sum_{i\neq j} J_{ij} \phi_i \phi_j\bigg\}.
\end{equation}
We first diagonalize $J \equiv U \Lambda U^T$, $\psi_i \equiv U^T_{ij} \phi_j$, so that
\begin{equation} \label{eq:linHam} \bham = \sum_{i=1}^N \int \ddx \left[ \frac{1}{2}\left(r + \lambda_i\right) \psi_i^2 + \frac{\kappa}{2} \left(\nabla \psi_i \right)^2\right].
\end{equation}
As $r$ is lowered, a linear instability is encountered at $r = - \lambda_1$, where $\lambda_1$ denotes the smallest eigenvalue of $J$ and $\psi_1$ is the corresponding eigenmode. Our interest is in the critical properties of the model at this transition. 

Here, we restrict $J$ to the GOE ensemble, where explicit expressions may be obtained for all integrals. The more general spectral edge results are provided in the main text. We take
\[\overline{J_{ij}} = 0, \qquad \overline{J_{ij} J_{k\ell}} = \left(\delta_{i k} \delta_{j \ell} +\delta_{i \ell} \delta_{j k} \right) \frac{\sigma^2}{N}.\]
As $N\rightarrow \infty$, the eigenvalues become distributed according to a Wigner semicircle
 \[\rho(\lambda) =\frac{1}{2\pi\sigma^2}\sqrt{4\sigma^2-\lambda^2}\,\mathbbm{1}\{|\lambda|\le 2\sigma\}.\]
 The large-$N$ critical point is thus located at $r_c = 2\sigma$, corresponding to a minimal eigenvalue $\lambda_1 \rightarrow -2\sigma$. We define the mass
 \[\tau \equiv r - r_c = r-2\sigma.\]
 
 \subsection{Heat capacity and upper critical dimension}
 For $\tau>0$, the saddle point solution is $\psi_i = 0$, and the Gaussian fluctuations about this state are described by the partition function
 \[Z=\prod_{i=1}^N\int\mathcal D\psi_i\,
\exp\left\{-\frac12\int \ddx\,\left[(r+\lambda_i)\psi_i^2+\kappa(\nabla\psi_i)^2\right]\right\}.\]
Evaluating the Gaussian integral in Fourier space, we obtain the free energy density per field
 \[\beta f \equiv -\frac{\log Z}{NV} = \frac{1}{N} \sum_{i=1}^N \frac{1}{2} \int \ddk \log(r+ \lambda_i + \kappa k^2)\]
up to additive constants. As $N\rightarrow \infty$, we may replace the sum over $i$ with an integral over the eigenvalue density $\frac{1}{N} \sum_{i=1}^N \rightarrow \int d\lambda \rho(\lambda)$:

\[\beta f = \frac{1}{4 \pi \sigma^2} \int \ddk \int_{-2\sigma}^{2\sigma} d\lambda \sqrt{4\sigma^2 -\lambda^2} \log(r + \lambda + \kappa k^2). \]
The integral over $\lambda$ can be written in terms of known quantities; differentiating once with respect to $r$, we have
\begin{equation} \label{eq:dfdr} \frac{\partial \beta f}{\partial r} = \frac{1}{2}\int \ddk \int \frac{d\lambda \rho(\lambda)}{r + \kappa k^2 +\lambda} = \frac{1}{2}\int \ddk\, g(r+\kappa k^2)\end{equation}
where $g(z)$ is the Stieltjes transform of the eigenvalue density~\cite{potters2020first}, equal to the normalized trace of the resolvent matrix $R(z) \equiv (z\mathbbm{1}-J)^{-1}$:
\[g(z) \equiv \lim_{N\rightarrow \infty}\frac{1}{N} \mathrm{Tr}\, R(z) = \int d\lambda\,\frac{\rho(\lambda)}{z-\lambda} = \frac{z- \sqrt{z^2 - 4\sigma^2}}{2\sigma^2}.\] 
Various quantities of interest can be computed from Eq.~\eqref{eq:dfdr}. For example, the heat capacity per field is found as
\[
C \equiv -\frac{\partial^2 \beta f}{\partial r^2}
= - \frac{1}{2}\int \ddk g'(r+\kappa k^2) = \int \ddk\, \frac{1}{4\sigma^2}\left[\frac{r+\kappa k^2}{\sqrt{(r+\kappa k^2)^2-4\sigma^2}}-1\right],
\]
 As $\tau\rightarrow 0^{+}$, the singular part is obtained by expanding the square root
\begin{equation} \label{eq:GaussianCsing} C^{\mathrm{sing}} \propto \int \ddk\, \frac{1}{\sqrt{\tau+\kappa k^2}} \propto \int^\Lambda_0 dk \frac{k^{d-1}}{\sqrt{\tau +\kappa k^2}}.\end{equation}
where $\Lambda$ is the ultraviolet (UV) cutoff. Notably, the integral is UV dominated for all $d>1$, and thus takes a finite cutoff-dependent value. This is unlike the standard case $\sigma = 0$, where we would have obtained an IR divergence $C^\mathrm{sing} \propto \tau^{-(4-d)/2}$ for $d<4$~\cite{kardar2007statistical}. As pointed out in Eq.~\eqref{main-eq:effectiveDimension} of the main text, this follows from the fact the singular part of the free energy has the same form as that of a single-field Gaussian theory in $d_\mathrm{eff} = d+3$.

\subsection{Critical correlation function}
The Fourier-space two-point correlation function of each eigenmode $\psi_i$ can be read from Eq.~\eqref{eq:linHam}:
\[
\langle \psi_i(\bk) \psi_j(\bk')\rangle
=
(2\pi)^d \delta(\bk+\bk') G_{ij}(\bk),
\qquad
G_{ij}(\bk)
\equiv
\frac{\delta_{ij}}{r+\lambda_i+\kappa k^2}
\]
The correlation functions of the original fields are then given by
\[
\langle \phi_i(\bk) \phi_j(\bk') \rangle
=
\sum_{a,b=1}^N U_{ia} U_{jb}
\langle \psi_a(\bk) \psi_b(\bk') \rangle
=
(2\pi)^d \delta(\bk+\bk')
\sum_{a=1}^N
\frac{U_{ia} U_{ja}}{r + \lambda_a + \kappa k^2}.
\]
The sum on the right hand side may be represented in matrix form as
\[
\sum_{a=1}^N
\frac{U_{ia} U_{ja}}{r + \lambda_a + \kappa k^2}
:=
U \left[(r+\kappa k^2) \mathbbm{1} + \Lambda \right]^{-1} U^T
=
\left[(r+\kappa k^2) \mathbbm{1} + J \right]^{-1}.
\]
The resolvent matrix has thus appeared again. It can be shown that for $r>r_c$, this object is self-averaging for large $N$~\cite{potters2020first}
\[
\left(z \mathbbm{1} + J \right)^{-1}
\rightarrow
\overline{\left(z \mathbbm{1} + J \right)^{-1}}
=
g(z) \mathbbm{1}
\]
We thus conclude
\begin{equation}
\label{eq:Sk}
\langle \phi_i(\bk) \phi_j(\bk') \rangle
\rightarrow
\delta_{ij} (2\pi)^d \delta(\bk + \bk') S(k),
\qquad
S(k)
=
\frac{2}{
r+\kappa k^2+
\sqrt{(r+\kappa k^2)^2-4\sigma^2}
}.
\end{equation}
Notably, the critical susceptibility is finite
\[
S(k=0)|_{r=2\sigma}
=
\frac{1}{\sigma}.
\]
There is, however, a cusp $\propto |k|$ at criticality, as can be seen by expanding in small $(\tau + \kappa k^2)$:
\begin{equation}
\label{eq:Skexpansion}
S(k)
=
\frac{1}{\sigma}
-\frac{\sqrt{\tau+\kappa k^2}}{\sigma^{3/2}}
+\frac{\tau+\kappa k^2}{2\sigma^2}
-\cdots .
\end{equation}
In real space, the constant above contributes a delta function, and the leading nontrivial behavior is captured by
\[
C(\bx)
\simeq
-\frac{1}{\sigma^{3/2}}
\int \ddk\,e^{i\bk\cdot\bx}
\sqrt{\tau + \kappa k^2}
=
-\frac{\sqrt{\kappa}}{\sigma^{3/2}}
\int \ddk\,e^{i\bk\cdot\bx}
\sqrt{k^2+\xi^{-2}},
\]
where $\xi \equiv \sqrt{\kappa/\tau}$. At criticality, dimensional analysis suggests that the cusp leads to a power law $\propto |x|^{-d-1}$. In fact, the finite-$\tau$ integral can be evaluated in terms of special functions using spherical symmetry: Choosing the polar axis along $\bx$, so that $\bk\cdot\bx=k|\bx|\cos\theta$, the angular integral is
\[
\int d\Omega_{d-1}\,e^{i\bk\cdot\bx}
=
(2\pi)^{d/2}
(k|\bx|)^{1-d/2}
\mathcal J_{d/2-1}(k|\bx|),
\]
where $\mathcal J_\nu$ is the Bessel function of the first kind. The remaining radial integral is one which can be found in Gradshteyn and Ryzhik, up to an analytic continuation which is valid for $|\bx|>0$:
\begin{align}
C(\bx)
&\simeq
-\frac{\sqrt{\kappa}}{\sigma^{3/2}}
\frac{|\bx|^{1-d/2}}{(2\pi)^{d/2}}
\int_0^\infty dk\,
k^{d/2}
\mathcal J_{d/2-1}(k|\bx|)
(k^2+\xi^{-2})^{1/2}
\nonumber \\
&=
\frac{\sqrt{\kappa}}{\sigma^{3/2}}
\frac{2^{(1-d)/2}}{\pi^{(d+1)/2}}
\frac{\xi^{-(d+1)/2}}{|\bx|^{(d+1)/2}}
K_{(d+1)/2}\left(\frac{|\bx|}{\xi}\right),
\qquad |\bx|>0,
\end{align}
where $K_\nu$ denotes the modified Bessel function of the second kind. The long distance behavior then follows from Bessel function asymptotics:
\[
C(\bx)
\simeq
\frac{\sqrt{\kappa}}{\sigma^{3/2}}
\times
\begin{cases}
\displaystyle
\frac{1}{(2\pi)^{d/2}}
\frac{\xi^{-d/2}}{|\bx|^{(d+2)/2}}
e^{-|\bx|/\xi},
& |\bx|\gg \xi,
\\[1em]
\displaystyle
\frac{\Gamma\!\left(\frac{d+1}{2}\right)}
{\pi^{(d+1)/2}}
\frac{1}{|\bx|^{d+1}},
& \Lambda^{-1} \ll |\bx| \ll \xi.
\end{cases}
\]
This confirms the critical power law $|\bx|^{-d-1}$ which was anticipated by dimensional analysis.

From the above results, we conclude the following values for the disordered-phase critical exponents in $d>1$:
\[
\alpha = 0,
\qquad
\gamma =0,
\qquad
\nu = \frac{1}{2},
\qquad
\eta=3.
\]
\subsection{Correlation function for general edge exponent $\theta$}
\label{sec:generalThetaCorrelation}
We now consider the general spectral-edge behavior defined in Eq.~\eqref{main-eq:edge} of the main text. From Eq.~\eqref{main-eq:structureFactor}, the leading nonanalytic part of the structure factor is
\begin{equation} \label{eq:thetaS}
S^{\mathrm{sing}}(k)
\simeq
-\frac{A\pi}{\sin(\pi\theta)}
\left(\tau+\kappa k^2\right)^\theta.
\end{equation}
The nontrivial part of the real space correlation function $C(\bx)$ is obtained from the inverse transform of Eq.~\eqref{eq:thetaS}; analytic terms in the full $S(k)$ contribute only contact terms to $C(\bx)$. Repeating the procedure used above in the Wigner case, and writing $\xi=\sqrt{\kappa/\tau}$, we obtain
\begin{equation}
\label{eq:generalThetaCorrelation}
C(\bx)
\simeq
\frac{2^{1+\theta}A\Gamma(1+\theta)\kappa^\theta}
{(2\pi)^{d/2}}
\frac{\xi^{-(d/2+\theta)}}{|\bx|^{d/2+\theta}}
K_{d/2+\theta}\left(\frac{|\bx|}{\xi}\right),
\qquad
|\bx|>0.
\end{equation}
Equivalently,
\[
C(\bx)
\sim
\frac{1}{|\bx|^{d+2\theta}}
f_\theta\left(\frac{|\bx|}{\xi}\right),
\qquad
f_\theta(y)\propto
y^{d/2+\theta}K_{d/2+\theta}(y).
\]
Provided $d>-2\theta$, Bessel function asymptotics show that $f_\theta(y)$ approaches a constant for $y\ll1$ and $f_\theta(y)\sim
y^{(d+2\theta-1)/2}e^{-y}$ for $y \gg 1$. Thus, at criticality,
\[
C(\bx)\sim |\bx|^{-(d+2\theta)},
\]
from which we obtain
\[
\nu=\frac{1}{2},
\qquad
\eta=2+2\theta.
\]
For $\theta=1/2$, these expressions recover the GOE results derived above.

\section{Component-wise $\phi^4$ model}
To study the stability of Gaussian exponents to nonlinearities, and to stabilize the low temperature phase, we add to each field a $\phi^4$ term,
\begin{equation}  \label{eq:siPhi4Ham} \bham = \int \ddx \left[\sum_{i=1}^N \left(\frac{r}{2}\phi_i^2 + \frac{\kappa}{2} \left(\nabla \phi_i\right)^2 + \frac{u}{4} \phi_i^4\right) + \frac{1}{2}\sum_{i\neq j} J_{ij} \phi_i \phi_j\right],
\end{equation}
where, for simplicity, we assume at the outset that $J_{ij}$ is a GOE random matrix, with independent components satisfying 
\[\overline{J_{ij}} = 0, \qquad \overline{J_{ij} J_{k\ell}} = \left(\delta_{i k} \delta_{j \ell} +\delta_{i \ell} \delta_{j k} \right) \frac{\sigma^2}{N}.\]
For $\sigma^2 = 0$, the model reduces to a collection of independent and identical $\phi^4$ models, whereas for $\kappa = 0$ it is a collection of independent soft spin Sherrington-Kirkpatrick models at each point in space. Here, we are interested in the critical properties of Eq.~\eqref{eq:siPhi4Ham} for $N \rightarrow \infty$ and $\sigma^2, \kappa > 0$.

\subsection{Replica analysis}
Unlike in the Gaussian case, working in the eigenbasis of $J$ is now cumbersome, as the $\phi^4$ interaction term in Eq.~\eqref{eq:siPhi4Ham} couples the eigenmodes nontrivially through the disordered tensor $\sum_a U_{ai}U_{aj}U_{ak}U_{a\ell}$. An alternative approach is provided by the replica method, which we pursue in this section, and which yields a result more amenable to perturbation theory. For $u=0$, the replica-symmetric high-temperature solution will recover the Gaussian results of Section~\ref{sec:gaussian} without relying on random matrix theory.

The replicated partition function is given by
\begin{align} Z^n &= \int \mathcal{D} \phi \, e^{-\sum_{\alpha=1}^n \bham[\phi_1^\alpha, \dots,  \phi_N^\alpha]} \nonumber \\
&= \int \mathcal{D} \phi \, \exp\left[-\int \ddx \sum_{i=1}^N\sum_{\alpha=1}^n \left(\frac{r}{2}(\phi_i^\alpha)^2 + \frac{\kappa}{2} \left(\nabla \phi_i^\alpha\right)^2 + \frac{u}{4} (\phi_i^\alpha)^4\right) - \frac{1}{2} \sum_{\alpha=1}^n \sum_{i\neq j} J_{ij} \int\ddx\, \phi_i^\alpha \phi_j^\alpha\right].
\end{align}
where we use the shorthand $\mathcal{D} \phi  \equiv \prod_{i, \alpha} \mathcal{D} \phi_i^\alpha$. Taking a quench average, we have
\[\overline{\exp\left[-\frac{1}{2} \sum_{\alpha=1}^n \sum_{i\neq j} J_{ij} \int \ddx \phi_i^\alpha \phi_j^\alpha \right]} = \exp\left[\frac{\sigma^2}{4N}\sum_{\alpha,\beta=1}^n\int_{\bx, \bx'} \,\sum_{i\neq j}\phi_i^\alpha(\bx)\phi_i^\beta(\bx')\phi_j^\alpha(\bx)\phi_j^\beta(\bx')\right].\]
We now define a physical overlap field,
\[\tilde{Q}_{\alpha \beta}(\bx, \bx') \equiv \frac{1}{N} \sum_{i=1}^N \phi_i^\alpha(\bx) \phi_i^\beta(\bx'),\]
and rewrite
\[\frac{1}{N}\sum_{i\neq j}\phi_i^\alpha(\bx)\phi_i^\beta(\bx')\phi_j^\alpha(\bx)\phi_j^\beta(\bx') = N \tilde{Q}_{\alpha\beta}^2(\bx, \bx') - \frac{1}{N} \sum_i { \big[ \phi_i^\alpha(\bx)\big]^2 \big[\phi_i^\beta(\bx')\big]^2}.
\]
The first term above is $\bigO{N}$, whereas the second is $\bigO{1}$ and may be neglected. The quenched-averaged replicated partition function is thus
\begin{align} \label{eq:znbar} \overline{Z^n} &= \int \mathcal{D} \phi \, \exp\left[- \sum_{i=1}^N\sum_{\alpha=1}^n  \int \ddx \left(\frac{r}{2}(\phi_i^\alpha)^2 + \frac{\kappa}{2} \left(\nabla \phi_i^\alpha\right)^2 + \frac{u}{4} (\phi_i^\alpha)^4\right) + 
 \frac{\sigma^2 N}{4} \sum_{\alpha, \beta=1}^n \int_{\bx, \bx'} \tilde{Q}_{\alpha \beta}^2(\bx, \bx')
\right].
\end{align}
We next introduce an independent Hubbard--Stratonovich field $Q_{\alpha\beta}(\bx,\bx')$:
\begin{align*}
&\exp\left[\frac{\sigma^2 N}{4}\sum_{\alpha,\beta}\int_{\bx,\bx'} \tilde{Q}_{\alpha\beta}^2(\bx,\bx')\right] \\
&\propto \int \mathcal D Q\,\exp\Bigg[
-\frac{\sigma^2 N}{4}\sum_{\alpha,\beta}\int_{\bx,\bx'} Q_{\alpha\beta}^2(\bx,\bx')
+\frac{\sigma^2 N}{2}\sum_{\alpha,\beta}\int_{\bx,\bx'} Q_{\alpha\beta}(\bx,\bx') \tilde{Q}_{\alpha\beta}(\bx,\bx')\Bigg] \\
&= \int \mathcal D Q\,\exp\Bigg[
-\frac{\sigma^2 N}{4}\sum_{\alpha,\beta}\int_{\bx,\bx'} Q_{\alpha\beta}^2(\bx,\bx') \Bigg]
\prod_i \exp\left[\frac{\sigma^2}{2}\sum_{\alpha,\beta}\int_{\bx,\bx'} Q_{\alpha\beta}(\bx,\bx')\phi_i^\alpha(\bx)\phi_i^\beta(\bx')\right].
\end{align*}
We see that the transformation has decoupled the fields. Equation~\eqref{eq:znbar} may thus be written
\[\overline{Z^n} = \int \mathcal D Q\,\exp\Bigg[
-\frac{\sigma^2 N}{4}\sum_{\alpha,\beta}\int_{\bx,\bx'} \,Q_{\alpha\beta}^2(\bx,\bx') \Bigg] \left[Z_1(Q)\right]^N \]
where
\begin{align} \label{eq:z1} Z_1(Q) \equiv \int \prod_{\alpha} \mathcal{D} \phi^\alpha \exp\Bigg[-\sum_\alpha \int \ddx \left(\frac{r}{2}(\phi^\alpha)^2 + \frac{\kappa}{2} \left(\nabla \phi^\alpha\right)^2 + \frac{u}{4} (\phi^\alpha)^4\right) \nonumber \\+ \frac{\sigma^2}{2} \sum_{\alpha \beta} \int_{\bx, \bx'} Q_{\alpha \beta} (\bx, \bx') \phi^\alpha(\bx) \phi^\beta(\bx')\Bigg].\end{align}
This has the form
\begin{equation} \label{eq:replicatedPartition} \overline{Z^n} = \int \mathcal{D} Q \, e^{-N \mathcal{S}[Q]},
\end{equation}
where the effective action is
\begin{equation} \label{eq:action} \mathcal{S}[Q] = \frac{\sigma^2}{4} \sum_{\alpha \beta} \int_{\bx,\bx'} Q^2_{\alpha \beta} - \log Z_1(Q). \end{equation}
For $N\rightarrow \infty$, the saddle point solution is given by
\[\frac{\delta \mathcal{S}}{\delta Q_{\alpha \beta}(\bx, \bx')} = 0\]
Evaluating the functional derivative gives a self-consistent equation for the Hubbard--Stratonovich field, which can be identified with the physical overlap in the large-$N$ limit:
\begin{equation} \label{eq:overlap} Q_{\alpha \beta}(\bx, \bx') = \langle \phi^\alpha (\bx) \phi^\beta(\bx') \rangle_{Z_1(Q)}\end{equation}
where $\langle \cdot \rangle_{Z_1(Q)}$ denotes an expectation value with respect to the single-site, many-replica measure defined by the partition function $Z_1(Q)$ given in Eq.~\eqref{eq:z1}. The effect of disorder averaging is thus the introduction of nonlocal quadratic interactions between and within replicas, with an interaction kernel $Q_{\alpha \beta}(\bx, \bx')$ determined self-consistently by the two-point correlation functions. 

By translational invariance, we may write the overlap as a single-variable function $Q_{\alpha \beta}(\bx, \bx') = Q_{\alpha \beta}(\bx -\bx')$. Commuting the product in Eq.~\eqref{eq:overlap} also yields the symmetry
\begin{equation} \label{eq:Qsymmetry} Q_{\alpha \beta}(\br) = Q_{\beta \alpha}(-\br) \end{equation}

\subsection{Mean field theory}
We first study Eqs.~\eqref{eq:replicatedPartition}-\eqref{eq:overlap} under a mean field approximation, where spatial fluctuations are neglected:
\begin{equation}
    Q_{\alpha\beta}(\bx, \bx') \equiv q_{\alpha \beta}, \qquad \phi^\alpha(\bx) \equiv m^\alpha.
\end{equation}
Defining the system volume $V\equiv \int \ddx$, the action becomes 
\begin{equation} \label{eq:MFaction} \mathcal{S} = \frac{\sigma^2 V^2}{4} \sum_{\alpha \beta} q_{\alpha \beta}^2 - \log Z_1,\end{equation}
where 
\begin{equation}\label{eq:z1MF}
Z_1(\{q_{\alpha \beta}\})\equiv \int \prod_{\alpha=1}^n dm^\alpha\,\exp\left[-V\sum_{\alpha=1}^n\left(\frac{r}{2}(m^\alpha)^2+\frac{u}{4}(m^\alpha)^4\right)+\frac{\sigma^2V^2}{2}\sum_{\alpha,\beta=1}^n q_{\alpha\beta}m^\alpha m^\beta\right],
\end{equation}
and Eq.~\eqref{eq:overlap} is unchanged at $q_{\alpha \beta} \equiv \langle m^\alpha m^\beta \rangle_{Z_1}$.

The system now has the structure of a soft-spin SK model, with fluctuations suppressed in the $V \rightarrow \infty$ limit~\cite{sherrington1975solvable}. We note, however, that since our model is defined with $r$ as the tunable parameter, $V$ should not be interpreted as the inverse temperature of a standard SK model; varying $r$ takes the system between a paramagnetic ($r>r_c$) and a spin-glass phase ($r<r_c$). Large $V$ suppresses fluctuations within the pure states of each phase, but $r_c$ takes a finite limit as $V\rightarrow \infty$. 

It is tempting to take the $V\rightarrow \infty$ limit at this stage and evaluate Eq.~\eqref{eq:z1MF} using the saddle point method. However, in the low temperature phase, this limit does not commute with the $n\rightarrow 0$ replica limit. We thus work at finite $V$, take $n \rightarrow 0$, and only then take $V\rightarrow \infty$.

\subsubsection{Replica-symmetric solution} \label{sec:RSSK}
We now make the replica-symmetric (RS) ansatz
\begin{equation}
    q^{\mathrm{RS}}_{\alpha \beta} = (p - q) \delta_{\alpha \beta} + q,
\end{equation}
which will allow us to locate the transition. Equation~\eqref{eq:z1MF} then reads,
\begin{equation}\label{eq:z1MFRS}
Z_1^{\mathrm{RS}}(p,q)=\int\prod_{\alpha=1}^n dm^\alpha\,\exp\left\{-V\sum_{\alpha=1}^n\left[\frac{r-\sigma^2V(p-q)}{2}(m^\alpha)^2+\frac{u}{4}(m^\alpha)^4\right]+\frac{\sigma^2V^2q}{2}\left(\sum_{\alpha=1}^n m^\alpha\right)^2\right\}.
\end{equation}
To decouple the replica integrals, we use introduce another auxiliary variable
\begin{equation}
\exp\left[\frac{\sigma^2V^2q}{2}\left(\sum_{\alpha=1}^n m^\alpha\right)^2\right]
=\int Dz\,\exp\left[\sigma V\sqrt{q}\,z\sum_{\alpha=1}^n m^\alpha\right],
\qquad Dz\equiv\frac{dz}{\sqrt{2\pi}}e^{-z^2/2}.
\end{equation}
We then have
\begin{equation}\label{eq:z1MFRSdecoupled}
Z_1^{\mathrm{RS}}(p,q)=\int Dz\,\left\{\int dm\,\exp\left[-V\left(\frac{r-\sigma^2V(p-q)}{2}m^2+\frac{u}{4}m^4-\sigma\sqrt{q}\,z\,m\right)\right]\right\}^n,
\end{equation}
with the action $\mathcal S^{\mathrm{RS}} = \frac{\sigma^2 V^2}{4} \left[np^2+n(n-1)q^2\right] - \log Z_1^{\mathrm{RS}}$. To take the $n\rightarrow 0$ replica limit, we use $\int Dz I^n = \int Dz\left[1+n \log I\right] + \mathcal{O}(n^2)$. The action is then expanded to leading order in $n$ as,
\begin{equation}\label{eq:smallnRSaction}
    \mathcal{S}^\mathrm{RS} = n\frac{\sigma^2 V^2}{4}\left( p^2 - q^2\right) - n\int Dz\,\log\left\{\int dm\,\exp\left[-V\left(\frac{r-\sigma^2V(p-q)}{2}m^2+\frac{u}{4}m^4-\sigma\sqrt q\,z\,m\right)\right]\right\} + \mathcal{O}(n^2)
\end{equation}
We now extremize with respect to $p$ and $q$. We first define the following fixed-$z$ average,
\begin{equation} \label{eq:zAverage}
\langle f(m)\rangle_z\equiv\frac{\int dm\,f(m)\,\exp\left\{-V\left[\frac{1}{2}(r-\sigma^2V(p-q))m^2+\frac{u}{4}m^4-\sigma\sqrt q\,z\,m\right]\right\}}{\int dm\,\exp\left\{-V\left[\frac{1}{2}(r-\sigma^2V(p-q))m^2+\frac{u}{4}m^4-\sigma\sqrt q\,z\,m\right]\right\}},
\end{equation}
in terms of which we find,
\begin{align} \label{eq:pderiv} \frac{\partial \mathcal{S}^{\mathrm{RS}}}{\partial p} &= \frac{n \sigma^2 V^2}{2}\left[p - \int Dz \langle m^2 \rangle_z\right] + \mathcal{O}(n^2), \\
\frac{\partial \mathcal{S}^{\mathrm{RS}}}{\partial q} &= \frac{n \sigma^2 V^2}{2}\left\{-q + \int D z \left[ \langle m^2 \rangle_z - \frac{1}{\sigma V \sqrt{q}} z \langle m \rangle_z \right]\right\} + \mathcal{O}(n^2). \label{eq:qderiv}
\end{align}
The $q$ derivative can be simplified using the identity $\int Dz\,  z F[z] = \int Dz F'[z]$, which gives.
\[
    \frac{1}{\sigma V\sqrt{q}} \int Dz z \langle m \rangle_z = \int Dz\left(\langle m^2 \rangle_z  - \langle m \rangle_z^2\right).
\]
Setting Eqs.~\eqref{eq:pderiv} and~\eqref{eq:qderiv} to zero then yields the self consistent equations
\begin{align}
    p &= \int Dz \langle m^2 \rangle_z, \\
    q &= \int Dz \langle m \rangle_z^2.
\end{align}
Finally, we may now take the $V \rightarrow \infty$ limit. In this limit, fluctuations in the fixed-$z$ measure of Eq.~\eqref{eq:zAverage} are suppressed, and $\langle m^2 \rangle_z \rightarrow \langle m \rangle_z^2$, implying that $p$ and $q$ are equal. However, the susceptibility
\[\chi \equiv V \left(\langle m^2 \rangle_{Z_1} - \langle m \rangle_{Z_1}^2 \right) = V(p-q)\]
which appears in Eq.~\eqref{eq:zAverage}, remains $\mathcal{O}(1)$; we will assume this to be true and then verify it self-consistently. For large $V$, the integral in Eq.~\eqref{eq:zAverage} is dominated by the vicinity of $m_z$, which satisfies the extremal condition
\begin{equation}\left[r-\sigma^2 \chi \right]m_z+um_z^3=\sigma\sqrt q\,z. \label{eq:mzdef}
\end{equation}
Writing $m = m_z + \delta m$ and expanding Eq.~\eqref{eq:zAverage} to Gaussian order yields the large-$V$ saddle point result
\begin{equation} \label{eq:fixedZvar}
\langle m^2\rangle_z-\langle m\rangle_z^2=\frac{1}{V\left[r-\sigma^2 \chi +3um_z^2\right]}+\mathcal{O}(V^{-2}),
\end{equation}
which verifies that $p - q \in \mathcal{O}(V^{-1})$. The large-$V$ self consistent system may then be written,
\begin{align}
\chi &= \int Dz\,\frac{1}{r-\sigma^2 \chi +3um_z^2} + \mathcal{O}(V^{-1}),  \label{eq:susceptibilityV} \\
q &= p = \int Dz\,m_z^2 + \mathcal{O}(V^{-1}).
\label{eq:qeqn}
\end{align}
This cannot in general be solved in closed form, but may be studied in the paramagnetic phase or in the glass phase just below the transition.

In the paramagnetic phase $q = p = 0$, Eq.~\eqref{eq:mzdef} sets $m_z = 0$. Equation~\eqref{eq:susceptibilityV} then yields
\begin{equation}
    \chi = \frac{r- \sqrt{r^2-4\sigma^2}}{2\sigma^2}, \qquad (r>r_c).
\end{equation}
This is real only above $r = 2\sigma$, indicating breakdown of the paramagnetic solution below
\begin{equation}
    r_c = 2\sigma.
\end{equation}
At $r=r_c$, the critical susceptibility is finite at $\chi_c = 1/\sigma$, as in the Gaussian model.
In the glass phase with $0 < (r_c - r)/r_c \ll 1$, $q$ and $m_z$ will be small. We may then solve Eq.~\eqref{eq:mzdef} perturbatively as
\begin{equation} \label{eq:mzexpansion}
m_z
=
\frac{\sigma\sqrt q\,z}{r-\sigma^2\chi}
-\frac{u\sigma^3q^{3/2}z^3}{(r-\sigma^2\chi)^4}
+\frac{3u^2\sigma^5q^{5/2}z^5}{(r-\sigma^2\chi)^7}
+\mathcal O(q^{7/2}).
\end{equation}
(only the first two terms are needed for what follows; the third will be used later to determine stability). Substituting into Eq.~\eqref{eq:qeqn} gives the self-consistent equation
\begin{equation}
q=\frac{\sigma^2q}{\left[r-\sigma^2 \chi\right]^2}-\frac{6u\sigma^4q^2}{\left[r-\sigma^2 \chi \right]^5}+\mathcal{O}(q^3).
\end{equation}
The nonzero solution is
\begin{align}
q &= \frac{\left(r-\sigma^2\chi\right)^3\left[\sigma^2-\left(r-\sigma^2\chi\right)^2\right]}{6u\sigma^4}+\mathcal{O}(q^2).
\end{align}
To expand this near $r = r_c$, we use the expansion of Eq.~\eqref{eq:susceptibilityV}
\begin{equation}
    \chi = \frac{1}{r - \sigma^2 \chi} - \frac{3 u q}{(r-\sigma^2 \chi)^2} + \mathcal{O}(q^2).
\end{equation}
With some algebra, this can be used to show 
\begin{equation} \label{eq:qofr}
    q = \frac{r_c - r}{3u} +\mathcal{O}(|r_c - r|^2).
\end{equation}
The off-diagonal overlap thus grows linearly below criticality.

\subsubsection{Instability of the replica-symmetric spin glass solution}
We now show that the RS solution derived in Sec.~\ref{sec:RSSK} is an unstable saddle of the action (Eq.~\ref{eq:MFaction}) in the glass phase $r<r_c$. The derivation follows that of de Almeida and Thouless (AT) for the standard SK model~\cite{de1978stability}. We first compute the Hessian 
\begin{equation}
    H_{\alpha \beta, \gamma \delta} = \frac{\partial^2 \mathcal{S}}{\partial q_{\gamma \delta} \partial q_{\alpha \beta}}.
\end{equation}
The derivatives are computed using
\begin{align}
\frac{\partial \mathcal S}{\partial q_{\alpha\beta}}
&=
\frac{\sigma^2V^2}{2}
\left[
q_{\alpha\beta}
-
\left\langle m^\alpha m^\beta\right\rangle
\right],
\\
\frac{\partial}{\partial q_{\gamma\delta}}
\left\langle m^\alpha m^\beta\right\rangle
&=
\frac{\sigma^2V^2}{2}
\left[
\left\langle
m^\alpha m^\beta m^\gamma m^\delta
\right\rangle
-
\left\langle m^\alpha m^\beta\right\rangle
\left\langle m^\gamma m^\delta\right\rangle
\right],
\end{align}
where the averages are with respect to $Z_1$, with the subscript dropped for brevity. The Hessian then reads
\begin{equation}
    H_{\alpha\beta,\gamma\delta}[\{q_{\alpha \beta}\}]
=
\frac{\sigma^2V^2}{2}
\delta_{\alpha\gamma}\delta_{\beta\delta}
-
\frac{\sigma^4V^4}{4}
\left[
\left\langle
m^\alpha m^\beta m^\gamma m^\delta
\right\rangle
-
q_{\alpha \beta}
q_{\gamma \delta}\right],
\end{equation}
where we have imposed the saddle point condition $q_{\alpha \beta} = \langle m^\alpha m^\beta\rangle$. To determine the stability of the RS solution, we ask whether $H^{\mathrm{RS}} \equiv H[q_{\alpha \beta}^\mathrm{RS}]$ has any unstable directions corresponding to RSB perturbations $q_{\alpha \beta} \equiv q^{\mathrm{RS}}_{\alpha \beta} + \eta_{\alpha \beta}$. Following the reasoning of Ref.~\cite{de1978stability}, the relevant perturbations for RSB are ``replicon" perturbations, which satisfy
\begin{equation} \label{eq:repliconeta}
    \eta_{\alpha \alpha} = 0, \qquad \eta_{\alpha \beta} = \eta_{\beta \alpha}, \qquad \sum_{\beta} \eta_{\alpha \beta} = 0.
\end{equation}
Multiplying such an $\eta$ by $H^{\mathrm{RS}}$ yields,
\begin{equation}
    \sum_{\gamma \delta} H^{\mathrm{RS}}_{\alpha \beta \gamma \delta} \eta_{\gamma \delta} = \frac{\sigma^2 V^2}{2} \eta_{\alpha \beta} - \frac{\sigma^4 V^4}{4} \left[\sum_{\gamma \delta}\langle m^\alpha m^\beta m^\gamma m^\delta \rangle_{\mathrm{RS}} \eta_{\gamma \delta} - q q_{\alpha \beta}^\mathrm{RS} \sum_{\gamma \delta} \eta_{\gamma \delta} \right].
\end{equation}
The final term vanishes by construction (Eq.~\ref{eq:repliconeta}). For the second term, we note that in the RS ansatz, all indices are interchangeable, and the value of the four-point correlator depends only on how many of the indices coincide. Terms with $\gamma = \delta$ not contribute as $\eta_{\gamma \gamma} = 0$. This leaves three other patterns: the pair $(\gamma, \delta)$ can (1) coincide with $(\alpha, \beta)$ or $(\beta, \alpha)$, or (2) share only one index with $(\alpha, \beta)$, or (3) share no indices. In case (1), the contributions to the sum are $2 \langle (m^\alpha)^2 (m^\beta)^2 \rangle_{\mathrm{RS}} \eta_{\alpha \beta}$. In case (2), the contribution is \(\left\langle
(m^\alpha)^2m^\beta m^\rho
\right\rangle_{\mathrm{RS}}
\sum_{\kappa\neq\alpha,\beta}
\left(
\eta_{\alpha\kappa}
+\eta_{\kappa\alpha}
+\eta_{\beta\kappa}
+\eta_{\kappa\beta}
\right)\), where $\rho$ is some arbitrary index different from $\alpha$ and $\beta$. Using Eq.~\eqref{eq:repliconeta}, this simplifies to $-4
\left\langle
(m^\alpha)^2m^\beta m^\rho
\right\rangle_{\mathrm{RS}}
\eta_{\alpha\beta}$. For case (3), similar arguments yields $2
\left\langle
m^\alpha m^\beta m^\rho m^\sigma
\right\rangle_{\mathrm{RS}}
\eta_{\alpha\beta}$, $\sigma$ distinct from $\alpha, \beta, \rho$. In summary, we see that $\eta$ is an eigenvector of $H^{\mathrm{RS}}$, 
\begin{equation}
    H^{\mathrm{RS}} \eta = \lambda_R \eta, \qquad \lambda_{R} = \frac{\sigma^2V^2}{2} - \frac{\sigma^4V^4}{2}\left[\left\langle(m^\alpha)^2(m^\beta)^2\right\rangle_{\mathrm{RS}} - 2\left\langle(m^\alpha)^2m^\beta m^\rho\right\rangle_{\mathrm{RS}} + \left\langle m^\alpha m^\beta m^\rho m^\sigma\right\rangle_{\mathrm{RS}}\right],
\end{equation}
with $\lambda_R$ identified as the replicon eigenvalue. To evaluate these expectation values, we use the decoupled form of $Z_1^{\mathrm{RS}}$ in Eq.~\eqref{eq:z1MFRSdecoupled}. At fixed $z$, distinct replicas are independent, and taking $n\rightarrow0$ gives
\begin{align}
\left\langle (m^\alpha)^2 (m^\beta)^2 \right\rangle_{\mathrm{RS}}
&=
\int Dz\,\langle m^2\rangle_z^2,
\\
\left\langle (m^\alpha)^2 m^\beta m^\rho \right\rangle_{\mathrm{RS}}
&=
\int Dz\,\langle m^2\rangle_z\langle m\rangle_z^2,
\\
\left\langle m^\alpha m^\beta m^\rho m^\sigma \right\rangle_{\mathrm{RS}}
&=
\int Dz\,\langle m\rangle_z^4.
\end{align}
The replicon eigenvalue then reads
\begin{equation}
\lambda_R
=
\frac{\sigma^2V^2}{2}
\left[
1-\sigma^2V^2
\int Dz\,
\left(
\langle m^2\rangle_z-\langle m\rangle_z^2
\right)^2
\right].
\end{equation}
Using the large-$V$ saddle point result Eq.~\eqref{eq:fixedZvar}, we have 
\begin{equation} \frac{2\lambda_R}{\sigma^2V^2}
=
1-\sigma^2\int Dz\,
\frac{1}{\left(r-\sigma^2\chi+3um_z^2\right)^2}
+O(V^{-1}).
\end{equation}
To determine the sign of this near the transition, we use the expansion of Eq.~\eqref{eq:mzexpansion}. For convenience, define $A \equiv r - \sigma^2 \chi$. Substituting Eq.~\eqref{eq:mzexpansion} into Eq.~\eqref{eq:qeqn}, retaining terms through order $q^3$, and dividing by the nonzero $q$ gives
\begin{equation} \label{eq:qExpansionStability}
1
=
\frac{\sigma^2}{A^2}
-\frac{6u\sigma^4q}{A^5}
+\frac{105u^2\sigma^6q^2}{A^8}
+\mathcal{O}(q^3),
\end{equation}
where we have used the Gaussian moments $\int Dz\,z^2=1$, $\int Dz\,z^4=3$, and $\int Dz\,z^6=15$. The integral appearing in the replicon eigenvalue can be expanded as
\begin{equation}
\frac{1}{(A+3um_z^2)^2}
=
\frac{1}{A^2}
-\frac{6u}{A^3}m_z^2
+\frac{27u^2}{A^4}m_z^4
+\mathcal{O}(q^3).
\end{equation}
Using Eq.~\eqref{eq:mzexpansion} and performing the Gaussian averages then gives
\begin{equation} \label{eq:repliconIntegralExpansion}
\sigma^2\int Dz\,\frac{1}{(A+3um_z^2)^2}
=
\frac{\sigma^2}{A^2}
-\frac{6u\sigma^4q}{A^5}
+\frac{117u^2\sigma^6q^2}{A^8}
+\mathcal{O}(q^3).
\end{equation}
Comparing Eqs.~\eqref{eq:qExpansionStability} and~\eqref{eq:repliconIntegralExpansion}, we obtain
\begin{equation}
\frac{2\lambda_R}{\sigma^2V^2}
=
-\frac{12u^2\sigma^6q^2}{A^8}
+\mathcal{O}(q^3)
+\mathcal{O}(V^{-1}).
\end{equation}
The leading term in Eq.~\eqref{eq:qExpansionStability} implies that $A=\sigma+\mathcal{O}(q)$. Taking $V\rightarrow\infty$ therefore yields
\begin{equation} \label{eq:repliconNearCritical}
\frac{2\lambda_R}{\sigma^2V^2}
=
-\frac{12u^2q^2}{\sigma^2}
+\mathcal{O}(q^3).
\end{equation}
Finally, using Eq.~\eqref{eq:qofr}, we find
\begin{equation}
\frac{2\lambda_R}{\sigma^2V^2}
=
-\frac{4}{3}
\left(\frac{r_c-r}{\sigma}\right)^2
+\mathcal{O}(|r_c-r|^3).
\end{equation}
Thus $\lambda_R<0$ immediately below $r_c$, showing that the RS solution is unstable upon entering the spin glass phase. In this phase, full Parisi RSB is expected to occur, by analogy to standard SK model.

\subsection{Perturbation theory in the high temperature phase}
We now return to the full the full $d$-dimensional model and characterize the high temperature paramagnetic phase, working perturbatively in powers of $u$. As in the mean field solution, expect a stable replica-symmetric solution with $Q_{\alpha \neq \beta} = 0$. The replicas are thus decoupled in Eq.~\eqref{eq:z1}, and the diagonal part follows
\begin{equation} \label{eq:highTselfCon} Q_{\alpha=\beta}(\bx, \bx') = \tilde{Q}_{\alpha=\beta}(\bx, \bx') = C(\bx-\bx') = \langle \phi(\bx) \phi(\bx') \rangle_{Z_1},
\end{equation}
where the expectation value is now evaluated with respect to a single-field Hamiltonian
\begin{equation} \label{eq:singleFieldHam} \bham_\mathrm{eff} = \int \ddx \left[\frac{r}{2} \phi^2 + \frac{\kappa}{2}(\nabla \phi)^2 + \frac{u}{4} \phi^4\right] - \frac{\sigma^2}{2} \int \ddx \ddx' C(\bx-\bx') \phi(\bx) \phi(\bx')
\end{equation}
The linear case $u=0$ is trivially solved in Fourier space; Equation~\eqref{eq:z1} becomes 
\[S_0(k) = \frac{1}{r+\kappa k^2 - \sigma^2 S_0(k)}.\]
This is Dyson's well-known formula for the resolvent, and its solution recovers exactly Eq.~\eqref{main-eq:wignersf} without reliance on random matrix theory results.

Unlike Eq.~\eqref{main-eq:fullHam}, Eq.~\eqref{eq:singleFieldHam} is directly amenable to diagrammatic perturbation theory; the main complication relative to standard $\phi^4$ theory is that the quadratic interaction kernel $S(k)$ must be determined self-consistently at each order. For example, the one-loop correction to the two-point function $\langle \phi(\bk) \phi(\bk') \rangle \equiv (2\pi)^d \delta(\bk+\bk') S(k)$ is given self-consistently by
\[S(k) = \frac{1}{r+\kappa k^2 -\sigma^2 S(k)} \left[1- 3 u \frac{1}{r+\kappa k^2 -\sigma^2 S(k)} \int \ddkp \frac{1}{r+\kappa k'^2 -\sigma^2 S_0(k')} + \bigO{u^2} \right].\]
It is helpful to consider the inverse
\[S(k)^{-1} = r +\kappa k^2 - \sigma^2 S(k) + 3 u \int \ddkp S_0(k') + \bigO{u^2},\]
where we have discarded $\mathcal{O}(u^2)$ terms. The effect of the nonlinearity can thus be interpreted as a shift in $r$:
\[r_\mathrm{eff} = r + 3 u \int \ddkp S_0(k'), \qquad S(k) = \frac{r_\mathrm{eff}+\kappa k^2- \sqrt{(r_\mathrm{eff}+\kappa k^2)^2 - 4\sigma^2}}{2\sigma^2}.\]
The shifted critical point is located as the point at which $S(k)$ develops the $|k|$ cusp and the correlation length diverges; namely, where $r_\mathrm{eff} = 2\sigma$:
\[r_c = 2\sigma - 3u \int \ddkp S_0(k')|_{r=2\sigma} + \bigO{u^2}.\]
The critical susceptibility thus remains finite and unrenormalized at $\bigO{u}$: $S(k=0)|_{r_\mathrm{eff} = 2\sigma} = 1/\sigma$. 

More generally, the $\bigO{u}$ corrections introduce only subleading singularities near criticality for $d>1$; for example, let us consider the fluctuation correction $\delta r(\tau_0) \equiv r_\mathrm{eff}(r) - r$ near the bare critical point $\tau_0 = r-2\sigma = 0$. We have
\begin{align*} \delta r(\tau_0) - \delta r(0) &= 3 u \int \ddk \left[ S_0(k; \tau_0) - S_0(k; 0) \right]\\
&\simeq - \frac{3 u}{\sigma^{3/2}} \int \ddk \left[\sqrt{\tau_0 + \kappa k^2} - \sqrt{\kappa} |k|\right] \propto \int_0^\Lambda dk k^{d-1} \left[\sqrt{\tau_0 + \kappa k^2} - \sqrt{\kappa} |k|\right]
\end{align*}
where we have used the expansion in Eq.~\eqref{main-eq:wignersf}. The integral can be made dimensionless by rescaling with the correlation length $\xi = \sqrt{\kappa/\tau_0}$:
\begin{align}
\delta r(\tau_0) - \delta r(0) \propto \xi^{-d-1} \int_0^{\Lambda \xi} dk\, k^{d-1} \left[\sqrt{1+k^2}-k\right] 
\end{align}
For $d>1$, the integral is UV divergent and determined by the cutoff. The analytic (in $\tau_0$) contribution comes from replacing the integrand with its large-$k$ form $k^{d-1}\left[\sqrt{1+k^2}-k \right] \sim k^{d-1}/2k$,
\begin{align}
\delta r(\tau_0) - \delta r(0) \propto \frac{\Lambda^{d-1}}{\kappa} \tau_0 + \text{nonanalytic correction} 
\end{align}
Subtracting the leading UV-singular part, the correction is $\xi^{-d-1} \int dk k^{d-1}\left[\sqrt{1+k^2}-k - 1/2k\right]$, which is IR-convergent in any $d>1$, and whose UV behavior is determined by $\xi^{-d-1} \int_0^{\Lambda \xi} dk k^{d-4}$. For $d<3$, the integral is finite, and the correction scales as $\xi^{-d-1} \propto \tau_0^{\frac{d+1}{2}}$, whereas for $d>3$ the integral is UV-divergent and the correction scales as $\xi^{-d-1} (\Lambda \xi)^{d-3} \propto \Lambda^{d-3} \tau_0^2/\kappa^2$. We thus have
\[\delta r (\tau_0) - \delta r (0) \propto \frac{\Lambda^{d-1}}{\kappa} \tau_0 + \mathcal{O}(\tau_0^{\frac{d+1}{2}}, \tau_0^2),\]
implying that no new leading singularities in $\tau_0$ are introduced by loop corrections near criticality for $d>1$.

\section{Spherical model}
To make further progress beyond the high temperature phase, we consider next a model whose low-temperature phase is simpler than the full RSB phase of Eq.~\eqref{eq:siPhi4Ham}. The modified model is described by the following Hamiltonian:
\begin{equation}  \label{eq:fsphericalHam} \bham = \int \ddx \left[\frac{r}{2}\phi_i^2 + \frac{\kappa}{2} \left(\nabla \phi_i\right)^2 + \frac{u}{4 N} (\phi_i^2)^2 + \frac{1}{2} J_{ij} \phi_i \phi_j\right],
\end{equation}
where summation over repeated indices is hereafter implied. The only change relative to Eq.~\eqref{eq:siPhi4Ham} is that the nonlinearity is now $\mathcal{O}(N)$-symmetric, so that for $J=0$, the vector $\vec{\phi} \equiv (\phi_1, \dots, \phi_N)$ becomes a spherical $\mathcal{O}(N)$ model as $N \rightarrow \infty$. The $J \neq 0$ model written above may then be regarded as a spherical model with random anisotropic mass. The $1/N$ factor in the nonlinear term is the typical choice for spherical models and is required to enforce that all terms are of equal order in $N$. 

The partition function then reads, in vector notation,
\[Z = \int \mathcal{D} \vec{\phi} \exp\left\{-\int \ddx \left[\frac{r}{2}\vec{\phi}^2 + \frac{\kappa}{2}(\nabla \vec{\phi})^2 + \frac{u}{4 N}(\vec{\phi}^2)^2 + \frac{1}{2} \vec{\phi}^T J \vec{\phi}\right]\right\}\]
Replicating and averaging over $J$ gives, to leading order in $N$,
\begin{align}
\overline{Z^n}\simeq\int\mathcal D\phi\,\exp\Bigg\{&-\frac12\sum_\alpha\int\ddx\left[r(\vec\phi^{\,\alpha})^2+\kappa(\nabla\vec\phi^{\,\alpha})^2\right] \nonumber\\
&+\frac{1}{4N}\sum_{\alpha\beta}\int\ddx\,d^d\bx'\left[\sigma^2-u\delta_{\alpha\beta}\delta(\bx-\bx')\right]
\left[\vec\phi^{\,\alpha}(\bx)\cdot\vec\phi^{\,\beta}(\bx')\right]^2\Bigg\}, \label{eq:sphericalReplicatedOneQ}
\end{align}
As before, the term in the disorder average with four fields carrying the same component index is subleading in $N$ and has been omitted. The term proportional to $\sigma^2$ is the remaining disorder contribution, while the local term proportional to $u$ is the original spherical nonlinearity. Thus, as in the zero-dimensional calculation, the two quartic terms may be decoupled together using a single overlap field:
\begin{align}
&\exp\left\{\frac{1}{4N}\sum_{\alpha\beta}\int\ddx\,d^d\bx'\left[\sigma^2-u\delta_{\alpha\beta}\delta(\bx-\bx')\right]
\left[\vec\phi^{\,\alpha}(\bx)\cdot\vec\phi^{\,\beta}(\bx')\right]^2\right\} \nonumber\\
&\quad\propto\int\mathcal DQ\,\exp\Bigg\{
-\frac N4\sum_{\alpha\beta}\int\ddx\,d^d\bx'
\left[\sigma^2-u\delta_{\alpha\beta}\delta(\bx-\bx')\right]Q_{\alpha\beta}^2(\bx,\bx') \nonumber\\
&\hspace{9em}+\frac12\sum_{\alpha\beta}\int\ddx\,d^d\bx'
\left[\sigma^2-u\delta_{\alpha\beta}\delta(\bx-\bx')\right]
Q_{\alpha\beta}(\bx,\bx')\vec\phi^{\,\alpha}(\bx)\cdot\vec\phi^{\,\beta}(\bx')\Bigg\}. \label{eq:sphericalOneQHS}
\end{align}
This yields 
\begin{equation}
\overline{Z^n}\propto\int\mathcal DQ\,e^{-N\mathcal{S}[Q]}, \label{eq:sphericalOneQZ}
\end{equation}
with
\begin{align}
\mathcal{S}[Q] &= \frac{1}{4}\sum_{\alpha\beta}\int\ddx\,d^d\bx'\left[\sigma^2-u\delta_{\alpha\beta}\delta(\bx-\bx')\right]Q_{\alpha\beta}^2(\bx,\bx')-\log Z_1(Q),\label{eq:sphericalAction} \\
Z_1(Q) &\equiv \int\prod_\alpha\mathcal D\phi^\alpha\exp\Bigg\{-\frac{1}{2}\sum_\alpha\int\ddx\left[r(\phi^\alpha)^2+\kappa(\nabla\phi^\alpha)^2\right] \nonumber\\
&\qquad+\frac{1}{2}\sum_{\alpha\beta}\int\ddx\,d^d\bx'\left[\sigma^2-u\delta_{\alpha\beta}\delta(\bx-\bx')\right]Q_{\alpha\beta}(\bx,\bx')\phi^\alpha(\bx)\phi^\beta(\bx')\Bigg\},
\end{align}
where it is emphasized that the $\phi^{\alpha}$ are now scalar variables rather than vectors, as the Hubbard field has decoupled the components. The saddle point equation is as before,
\begin{equation} \label{eq:sphericalSelfCon}
Q_{\alpha\beta}(\bx,\bx') = \left\langle\phi^\alpha(\bx)\phi^\beta(\bx')\right\rangle_{Z_1(Q)}.
\end{equation}
What distinguishes this from the SK case is $Z_1$ can be evaluated explicitly in terms of $Q$ because it is a Gaussian integral. Let us first impose translational invariance $Q_{\alpha \beta}(\bx, \bx') = Q_{\alpha \beta}(\bx - \bx')$, so that $Z_1$ may be expressed in Fourier space:
\begin{align} \label{eq:Z1Spherical}
    Z_1(Q) &\equiv \int\prod_\alpha\mathcal D\phi^\alpha\exp\Bigg\{-\frac{1}{2}\sum_\alpha \int \ddk \left[(r + u p_\alpha +\kappa k^2) |\phi^\alpha(\bk)|^2\right] \nonumber\\
&\qquad+\frac{\sigma^2}{2}\sum_{\alpha\beta}\int \ddk Q_{\alpha\beta}(\bk)\phi^\alpha(\bk)\phi^\beta(-\bk)\Bigg\},
\end{align}
where 
\begin{equation} \label{eq:pdef} p_\alpha \equiv Q_{\alpha \alpha}(\bx-\bx'=0) = \int \ddk Q_{\alpha \alpha}(\bk).
\end{equation}
Equation~\eqref{eq:sphericalSelfCon} may now be written in Fourier space as
\[Q(\bk) = \left[(r+ \kappa k^2) \mathbbm{1} + u\,  \mathrm{diag}(p_1, \dots, p_n) - \sigma^2 Q(\bk)\right]^{-1}.\]
It is helpful to rewrite this as
\begin{equation} \label{eq:sphericalSelfConDyson} Q^{-1}(\bk) + \sigma^2 Q(\bk) = \left(r + \kappa k^2\right) \mathbbm{1} + u \mathrm{diag}(p_1, \dots, p_n). \end{equation}
Since the left hand side commutes with $Q$, we must have $\left[Q, \mathrm{diag}(p_1, \dots, p_n)\right] = 0$, which implies
\[
(p_\alpha-p_\beta)Q_{\alpha\beta}(\bk)=0.
\]
Thus, only replica pairs $(\alpha, \beta)$ with $p_\alpha = p_\beta$ can have nonzero overlap $Q_{\alpha \beta}$. Let us now make the RS ansatz
\begin{align*}
    Q_{\alpha \alpha}(\bk) &\equiv S(k), \\
    Q_{\alpha \neq \beta}(\bk) &\equiv q(k),\\
    p_\alpha &\equiv p,
\end{align*}
where we have assumed rotational invariance in space. Substituting this form into Eq.~\eqref{eq:sphericalSelfConDyson} yields, after some rearrangement,
\begin{align}
\frac{1}{S(k)-q(k)}+\sigma^2\left[S(k)-q(k)\right] &= r+u p+\kappa k^2,\label{eq:SkSphericalSelfCon} \\
q(k)\left\{\sigma^2-\frac{1}{\left[S(k)-q(k)\right]\left[S(k)+(n-1)q(k)\right]}\right\} &= 0. \label{eq:sphericalSelfCon2}
\end{align}

\subsection{High temperature phase}
The high temperature phase corresponds to $q(k) = 0$. In that case, Eq.~\eqref{eq:SkSphericalSelfCon} recovers the standard scalar Dyson equation, whose solution is
\begin{equation} \label{eq:HTsphericalSF}
S_\mathrm{HT}(k)=\frac{2}{r+u p+\kappa k^2+\sqrt{\left(r+u p+\kappa k^2\right)^2-4\sigma^2}}
\end{equation}
(there is a second solution which is discarded because it diverges as $\sigma^2\rightarrow 0$). This has the same structure as the Gaussian theory explored earlier, except that it depends on the self-consistent parameter $p(r)$. In the spherical ferromagnet, it is the singularities of $p(r)$ as $r\rightarrow r_c$ which determine the upper and lower critical dimensions. Here, we have
\begin{equation}p = \int \ddk \frac{2}{r+u p+\kappa k^2+\sqrt{\left(r+u p+\kappa k^2\right)^2-4\sigma^2}}.  \label{eq:HTp}\end{equation}
The structure of the theory is similar to that of the $\phi^4$ case at one loop level: the nonlinearity shifts the mass as $r \rightarrow r + u p$, with $p$ given as an integral of the Gaussian structure factor. As before, the critical point is identified as the point where $S(k)$ develops its cusp; namely, where $r_c = 2\sigma - u p$. There, we have
\begin{equation} p(r_c) = \int \ddk S_c(k) \equiv p_c, \qquad
S_c(k) \equiv \frac{2}{2\sigma + \kappa k^2 + \sqrt{(2\sigma + \kappa k^2)^2 - 4\sigma^2}}, \label{eq:pcritical} \end{equation}
where the integral is understood up to the UV cutoff $\Lambda$. $p_c$ is always IR finite, as the integrand approaches $1/\sigma$ for small $k$. This is unlike in the spherical ferromagnet, where the divergence of $p_c$ in $d \leq 2$ implies a lower critical dimension of 2. To determine whether the present model supports long ranged order in $d=1$, however, would require a proper domain wall analysis.

The shifted mass $\tau_\mathrm{eff} \equiv r + u p(r) - 2 \sigma$ vanishes linearly with $(r-r_c)$, so that the Gaussian exponents apply. To see this, we first compute $p'(r_c)$:
\begin{align} 
p'(r_c) &= -2 \int \ddk \left.\frac{1 + u p'(r)}{\sqrt{\left(r + u p + \kappa k^2\right)^2 - 4\sigma^2} \left(r + u p + \kappa k^2 + \sqrt{\left(r + u p + \kappa k^2\right)^2 - 4\sigma^2}\right)}\right|_{r = r_c}. \nonumber \\
&= \frac{1 + u p'(r_c)}{2\sigma^2} \int \ddk \left( 1 - \frac{2\sigma + \kappa k^2}{|k|\sqrt{\kappa(4\sigma + \kappa k^2)}} \right) \label{eq:pprime}
\end{align}
Solving for $p'(r_c)$ yields
\begin{equation} \label{eq:pprimeSimple} p'(r_c) = \frac{\int \ddk \left( 1 - \frac{2\sigma + \kappa k^2}{|k|\sqrt{\kappa(4\sigma + \kappa k^2)}} \right)}{2\sigma^2 - u \int \ddk \left( 1 - \frac{2\sigma + \kappa k^2}{|k|\sqrt{\kappa(4\sigma + \kappa k^2)}} \right)}\end{equation}
Since $\tau_\mathrm{eff}'(r) = 1+up'(r_c)$, we have
\begin{equation}\tau_{\mathrm{eff}}'(r_c)
=\left[
1
-\frac{u}{2\sigma^2}\int\ddk
\left(
1-
\frac{2\sigma+\kappa k^2}
{|k|\sqrt{\kappa(4\sigma+\kappa k^2)}}
\right)
\right]^{-1}.\end{equation}
At small $k$, the integrand behaves as $|k|^{-1}$, so the infrared contribution is proportional to $\int_0^\Lambda dk k^{d-2}$. This is finite in $d>1$. It can verified that the integral is strictly negative, $0<\tau_{\mathrm{eff}}'(r) < 1$. We thus have a standard expansion near criticality,
\begin{equation}
    \tau_{\mathrm{eff}}(r) \simeq \tau_{\mathrm{eff}}'(r) (r-r_c)
\end{equation}
implying that the nonlinearity does not renormalize Gaussian exponents in $d>1$. 
\subsection{Low temperature phase}
In the low temperature phase, the off diagonal overlap $q$ is nonzero. We will first prove that the off-diagonal overlap must be uniform in space: Suppose $q(k) \neq 0$ for some $k$. Eq.~\eqref{eq:sphericalSelfCon2} then gives
\[\frac{1}{S(k)-q(k)} = \sigma^2 \left[S(k) + (n-1) q(k)\right]\]
In the $n\rightarrow 0$ replica limit, this reads
\begin{equation} \left[S(k) - q(k)\right]^2 = \frac{1}{\sigma^2}. \label{eq:lowtempConnected}
\end{equation}
To choose the correct root, we note that $S(k) - q(k)$ is required to be positive: To see this, consider
\[
\left\langle \left[\phi^\alpha(\bk)-\phi^\beta(\bk)\right]\left[\phi^\alpha(\bk')-\phi^\beta(\bk')\right]\right\rangle=2(2\pi)^d\delta(\bk+\bk')\left[S(k)-q(k)\right].
\]
Setting $\bk'=-\bk$, the left-hand side becomes $\left\langle\left|\phi^\alpha(\bk)-\phi^\beta(\bk)\right|^2\right\rangle\geq0$. Since $(2\pi)^d\delta(\mathbf 0)$ is a positive volume factor, it follows that
\[
S(k)-q(k)\geq0.
\]
We thus take the positive square root in Eq.~\eqref{eq:lowtempConnected}: $S(k) - q(k) = 1/\sigma$. Substituting into Eq.~\eqref{eq:SkSphericalSelfCon} then leads us to conclude that, for any $k$ for which $q(k) \neq 0$,
\begin{equation}\label{eq:qnonzeroCondition} r + u p + \kappa k^2 = 2\sigma.
\end{equation}
This cannot hold simultaneously for all $k$, implying that $q(k)$ can be nonzero only for one value of $k^2$. To then see that only the uniform $q(k=0)$ can be nonzero, we note the inequality $x^{-1} + \sigma^2 x \geq 2\sigma$ for $x>0$, which allows us to write
\begin{equation} \label{eq:qinequality} r + u p + \kappa k^2 = \frac{1}{S(k) - q(k)} + \sigma^2\left[S(k)-q(k)\right] \geq 2\sigma
\end{equation}
for all $k$. But if Eq.~\eqref{eq:qnonzeroCondition} were to hold for any nonzero $k$, then this would contradict Eq.~\eqref{eq:qinequality} with $k$ set to zero. We thus conclude that only $q(k = 0)$ may be nonzero, implying that the off diagonal overlap remains uniform in space. Setting $k=0$ in Eq.~\eqref{eq:qnonzeroCondition} then shows that, in the low temperature phase, $p$ is pinned to the value,
\begin{equation} \label{eq:pinnedP} p = \frac{2\sigma - r}{u}\end{equation}
Let us then write
\[q(k) = (2\pi)^d q \delta^{(d)}(\bk).\]
Using the definition of $p$ in Eq.~\eqref{eq:pdef}, we have
\begin{equation} \label{eq:pminusq} p - q = \int \ddk \left[S(k)-q(k)\right]\end{equation}
The integrand $\left[S(k)-q(k)\right]$ satisfies Eq.~\eqref{eq:SkSphericalSelfCon}, which for $p$ as in Eq.~\eqref{eq:pinnedP} reads
\[\frac{1}{S(k)-q(k)} + \sigma^2 \left[S(k)-q(k)\right] = 2\sigma + \kappa k^2.\]
But this is precisely the equation satisfies by the critical structure factor $S_c$ obtained from the $q=0$ high temperature side (Eq.~\ref{eq:pcritical}). The integral on the RHS of Eq.~\eqref{eq:pminusq} is thus identical to that in Eq.~\eqref{eq:pcritical}, leading us to conclude
\[p - q = p_c.\]
In terms of $r$, this reads
\[q = \frac{r_c - r}{u}.\]
The overlap thus vanishes linearly near criticality, as in the mean-field case, implying that the exponent for the vanishing of the order parameter is not renormalized by fluctuations in this spherical model.

In the ordered phase, the structure factor assumes its critical form, except for the shift in the uniform mode:
\[S(k) = S_c(k) + q(k) = \frac{2}{2\sigma + \kappa k^2 + \sqrt{(2\sigma + \kappa k^2)^2 - 4\sigma^2}} + (2\pi)^d q \delta^{(d)}(\bk).\]
In real space, the correlation function $\langle \phi^{\alpha}(\bx) \phi(\bx') \rangle$ thus decays to a constant $q$, implying the emergence of long-ranged order below $r_c$. The connected component is given by the inverse transform of $S_c(k)$, and thus has a diverging correlation length but a finite susceptibility. The $\nu$ exponent therefore has no sensible ordered-side definition.

It is interesting that this model displays generic scale invariance all throughout the ordered phase, despite the absence of a Goldstone mode (the quenched random coupling breaks $O(N)$ symmetry). 

\subsection{Quench-averaged free energy and heat capacity}
Having computed the overlap matrix, we may now evaluate the quench-averaged free energy using the replica trick. At the large-$N$ saddle point, Eq.~\eqref{eq:sphericalOneQZ} gives
\begin{equation}
    \overline{Z^n} \simeq e^{- N \mathcal{S}[Q]}
\end{equation}
with $Q$ taking its RS saddle point value. Let us start with the first term in $\mathcal{S}[Q]$ from Eq.~\eqref{eq:sphericalAction}:
\begin{equation}\frac{1}{4}\sum_{\alpha\beta}\int_{\bx, \bx'} \left[\sigma^2-u\delta_{\alpha\beta}\delta(\bx-\bx')\right]Q_{\alpha\beta}^2(\bx - \bx') = \frac{V}{4} \left[\sigma^2 \sum_{\alpha \beta} \int d^d\mathbf{r} \, Q_{\alpha \beta}^2(\mathbf{r}) - u\sum_\alpha Q_{\alpha \alpha}^2(0)\right] \label{eq:ActionBigPart1} 
\end{equation}
Equation~\eqref{eq:pdef} gives us $\sum_\alpha Q_{\alpha \alpha}^2(0) = n p^2$. For the other term, the off-diagonal contribution is uniform $Q_{\alpha\neq\beta}^2 = q^2$. The diagonal contribution is nonuniform but decays to a constant $q$. We separate out the uniform parts and write what remains as a Fourier integral
\begin{align}
\sum_{\alpha\beta}\int d^d\mathbf{r}\,Q_{\alpha\beta}^2(\br)
&=Vn(n-1)q^2+n\int d^d\mathbf{r}\,Q_{\alpha\alpha}^2(\br) \nonumber\\
&=Vn^2q^2+n\int d^d\mathbf{r}\left\{\left[Q_{\alpha\alpha}(\mathbf{r})-q\right]^2+2q\left[Q_{\alpha\alpha}(\mathbf{r})-q\right]\right\} \nonumber\\
&=Vn^2q^2+n\int\ddk\,S_{\mathrm{conn}}^2(k)+2nqS_{\mathrm{conn}}(k=0). \label{eq:actionPt1}
\end{align}
where $S_{\mathrm{conn}}(k)$ corresponds to the Fourier transform of the connected correlation function, obtained from the full structure factor by subtracting the uniform contribution:
\begin{equation}
    S_{\mathrm{conn}}(k) \equiv S(k) - (2\pi)^d q \delta^{(d)}(\bk) = \begin{cases}
        S_{\mathrm{HT}}(k), &r>r_c, \\
        S_c(k), &r<r_c.
    \end{cases}
\end{equation}
where $S_{\mathrm{HT}}(k)$ and $S_c$ are respectively the high temperature (Eq.~\ref{eq:HTsphericalSF}) and critical (Eq.~\ref{eq:pcritical}) structure factors. 

Together, Eqs.~\eqref{eq:ActionBigPart1} and~\eqref{eq:actionPt1} let us write the saddle point action (Eq.~\eqref{eq:sphericalAction}) as
\begin{equation} \label{eq:actionSphericalPartiallyEval}
\mathcal{S}[Q]=\frac{Vn}{4}\left[\sigma^2\int\ddk\,S_{\mathrm{conn}}^2(k)+2\sigma^2qS_{\mathrm{conn}}(k=0)-u p^2\right]-\log Z_1(Q)+\mathcal O(n^2).
\end{equation}
It remains to evaluate $\log Z_1(Q)$. Let us separate the uniform zero mode as
\begin{align} Q_{\alpha \beta}(\bk) &= S_\mathrm{conn}(k) \delta_{\alpha \beta} + q (2\pi)^d  \delta^{(d)}(\bk), \label{eq:QRSAnsatz}
\end{align}
The full partition function may thus be written
\begin{equation}
Z_1(Q) = \int {\textstyle\prod_\alpha}\mathcal D\phi^\alpha\exp\Bigg\{-\frac{1}{2} \sum_{\alpha \beta} \int \ddk \phi^{\alpha}(-\bk) \mathcal{M}_{\alpha \beta}(\bk) \phi^\beta(\bk) \Bigg\},
\end{equation}
where
\begin{equation}
\mathcal{M}_{\alpha\beta}(\bk)
=\left[r+u p+\kappa k^2-\sigma^2S_{\mathrm{conn}}(k)\right]\delta_{\alpha\beta}-\sigma^2q(2\pi)^d\delta^{(d)}(\bk).
\end{equation}
Since this is diagonal in $\bk$ space, the Gaussian integral is given by
\begin{equation} \label{eq:Z1logdet} -\log Z_1(Q) = \frac{V}{2} \int \ddk \log \det \mathcal M(\bk), \end{equation}
up to additive constants, where the determinant is over replica indices. The matrix $\mathcal{M}_{\alpha\beta}(\bk)$ has $(n-1)$ degenerate eigenvectors orthogonal to $(1,\dots, 1)$, and a distinct eigenvalue corresponding to $(1,\dots, 1)$.
\begin{align}
\log\det\mathcal M(\bk)
={}&(n-1)\log\left[r+u p+\kappa k^2-\sigma^2S_{\mathrm{conn}}(k)\right] \nonumber\\
&+\log\left[r+u p+\kappa k^2-\sigma^2S_{\mathrm{conn}}(k)-n\sigma^2q(2\pi)^d\delta^{(d)}(\bk)\right] \nonumber\\
={}&n\log\left[r+u p+\kappa k^2-\sigma^2S_{\mathrm{conn}}(k)\right] - \frac{n\sigma^2q(2\pi)^d\delta^{(d)}(\bk)}{r+u p+\kappa k^2-\sigma^2S_{\mathrm{conn}}(k)} + \mathcal{O}(n^2).
\end{align}
Substituting back into Eq.~\eqref{eq:Z1logdet} and then Eq.~\eqref{eq:actionSphericalPartiallyEval} yields
\begin{align}
\mathcal{S}[Q]
=\frac{Vn}{4}\Bigg\{&
\int\ddk\left[\sigma^2S_{\mathrm{conn}}^2(k)+2\log\left(r+u p+\kappa k^2-\sigma^2S_{\mathrm{conn}}(k)\right)\right] \nonumber\\
&+2\sigma^2qS_{\mathrm{conn}}(k=0)-u p^2-\frac{2\sigma^2q}{r+u p-\sigma^2S_{\mathrm{conn}}(k=0)}\Bigg\}+\mathcal O(n^2).
\end{align}
This can be simplified considerably using the saddle point equations: Equation~\ref{eq:SkSphericalSelfCon} gives $r+u p+\kappa k^2-\sigma^2S_{\mathrm{conn}}(k) = 1/S_{\mathrm{conn}}(k)$. Furthermore, the terms with $S_\mathrm{conn}(k=0)$ vanish at leading order
\begin{equation} 2\sigma^2qS_{\mathrm{conn}}(k=0)-\frac{2\sigma^2q}{r+u p-\sigma^2S_{\mathrm{conn}}(k=0)} = 0 + \mathcal{O}(n).
\end{equation}
This is obviously true in the high temperature phase, where $q=0$. In the low temperature phase, Eq.~\eqref{eq:sphericalSelfCon2} gives $S_\mathrm{conn}(\bk=0) = 1/\sigma + \mathcal{O}(n)$ and, by Eq.~\eqref{eq:pinnedP}, $r + u p = 2\sigma$, so the two terms cancel at leading order. The action thus reduces to the following form, valid in both phases:
\begin{equation}
\mathcal{S}[Q]=\frac{Vn}{4}\left\{\int\ddk\left[\sigma^2S_{\mathrm{conn}}^2(k)-2\log S_{\mathrm{conn}}(k)\right]-u p^2\right\}+\mathcal O(n^2).
\end{equation}
At last, we invoke the replica trick:
\begin{equation}
    \overline{\log Z} = \lim_{n\rightarrow 0} \frac{\overline{Z^n}-1}{n} = \lim_{n\rightarrow 0} \frac{e^{-N \mathcal{S}[Q]}-1}{n}.
\end{equation}
Since $\mathcal{S}[Q] \in \mathcal{O}(n)$, we obtain the following expression for the quench-averaged free energy density per field:
\begin{equation}
    \beta f \equiv -\frac{\overline{\log Z}}{NV} = \frac{1}{4}\left\{\int\ddk\left[\sigma^2S_{\mathrm{conn}}^2(k)-2\log S_{\mathrm{conn}}(k)\right]-u p^2\right\}.
\end{equation}
In the ordered phase, this has a particularly simple $r$ dependence as $S_\mathrm{conn}$ is pinned to the critical $S_c$, with no $r$ dependence, so that 
\begin{equation}
\beta f(r)
=
\beta f(r_c)
+\frac{p_c}{2}(r-r_c)
-\frac{(r-r_c)^2}{4u}, \qquad r<r_c.
\end{equation}
The heat capacity is then constant in this phase
\begin{equation} \label{eq:lowTHeatcap}
    -\frac{\partial^2 \beta f(r)}{\partial r^2} = \frac{1}{2u}, \qquad r<r_c.
\end{equation}
In the high temperature phase, on the other hand, we have
\begin{equation}
\frac{\partial\beta f}{\partial r}
=
\frac{1}{2}\int\ddk\left[\sigma^2S_{\mathrm{HT}}(k)-\frac{1}{S_{\mathrm{HT}}(k)}\right]\frac{\partial S_{\mathrm{HT}}(k)}{\partial r}
-\frac{u}{2}p p'(r).
\end{equation}
Differentiating Eq.~\eqref{eq:SkSphericalSelfCon} with $q(k)=0$ gives
\begin{equation}
\left[\sigma^2-\frac{1}{S_{\mathrm{HT}}^2(k)}\right]
\frac{\partial S_{\mathrm{HT}}(k)}{\partial r}
=1+u p'(r),
\end{equation}
so that
\begin{equation}
\frac{\partial\beta f}{\partial r}
=
\frac{1+u p'(r)}{2}\int\ddk\,S_{\mathrm{HT}}(k)
-\frac{u}{2}p p'(r).
\end{equation}
But the integral above is the definition of $p$, implying that the terms proportional to $p'$ cancel. The high temperature heat capacity then reads
\begin{equation} -\frac{\partial^2 \beta f(r)}{\partial r^2} = -\frac{p'(r)}{2} 
\end{equation}
with $p$ obtained taking its high temperature form in Eq~\eqref{eq:HTp}. On approaching the critical point from the high temperature side, the limiting heat capacity is
\begin{equation} \label{eq:highTcriticalC} -\frac{\partial^2 \beta f(r)}{\partial r^2} = -\frac{p'(r)}{2} =
\frac{1}{2u + 4\sigma^2/B}, \qquad r \rightarrow r_c^+
\end{equation}
where
\begin{equation} B = \int \ddk \left( \frac{2\sigma + \kappa k^2}{|k|\sqrt{\kappa(4\sigma + \kappa k^2)}} -1\right)\end{equation}
which follows from the self-consistent solution for $p'(r_c)$ obtained in Eq.~\eqref{eq:pprimeSimple}. As long as $B$ is finite, Eq.~\eqref{eq:highTcriticalC} is different from Eq.~\eqref{eq:lowTHeatcap}, implying that the heat capacity is discontinuous at the critical point. This is the case in any $d>1$. In $d=1$, $B$ is logarithmically IR divergent, so that Eq.~\eqref{eq:highTcriticalC} approaches the low temperature value $1/2u$.

In $d=0$, the heat capacity has a cusp but is continuous at the critical point. This is also the case for the standard SK model. Our results show that in the $d>0$ model, spatial fluctuations renormalize the continuous cusp into a discontinuous jump. It can be verified that $B$ is always positive, so that the jump corresponds to an increase in heat capacity as $r$ is lowered.

\subsection{Stability of the replica-symmetric solution}
Here, we verify the stability of the replica-symmetric saddle solved above. To do so, we consider the Hessian of $\mathcal{S}$:
\begin{align} H_{\alpha \beta, \gamma \delta}(\bx_1, \bx_2, \bx_3, \bx_4) &= \frac{\delta^2 \mathcal{S}}{\delta Q_{\alpha \beta}(\bx_1, \bx_2) \delta Q_{\gamma \delta}(\bx_3, \bx_4)}.
\end{align}
It is useful to define the kernel
\begin{equation}
    K_{\alpha \beta}(\bx, \bx') \equiv \sigma^2 - u \delta_{\alpha \beta} \delta(\bx-\bx'),
\end{equation}
so that the action reads
\begin{equation}
\mathcal{S}[Q]
=
\frac{1}{4}\sum_{\alpha\beta}
\int d^d\bx\,d^d\bx'\,
K_{\alpha\beta}(\bx,\bx')
Q_{\alpha\beta}^2(\bx,\bx')
-\log Z_1(Q).
\end{equation}
The first derivative, previously used to obtain the saddle point equations, is
\begin{equation} \label{eq:actionFirstDeriv}
    \frac{\delta \mathcal{S}[Q]}{\delta Q_{\gamma \delta} (\bx_3, \bx_4)} = \frac{1}{2} K_{\gamma \delta}(\bx_3, \bx_4) \left[Q_{\gamma \delta}(\bx_3, \bx_4) - \left \langle \phi^\gamma(\bx_3) \phi^{\delta} (\bx_4) \right \rangle_{Z_1(Q)}\right],
\end{equation}
with no summation implied. To compute the second derivative, we use
\begin{align}
\frac{\delta}{\delta Q_{\alpha\beta}(\bx_1,\bx_2)}\left\langle\phi^\gamma(\bx_3)\phi^\delta(\bx_4)\right\rangle_{Z_1(Q)}
&=\frac{1}{2}K_{\alpha\beta}(\bx_1,\bx_2)\Big[\left\langle\phi^\gamma(\bx_3)\phi^\delta(\bx_4)\phi^\alpha(\bx_1)\phi^\beta(\bx_2)\right\rangle_{Z_1(Q)} \nonumber\\
&\qquad-\left\langle\phi^\gamma(\bx_3)\phi^\delta(\bx_4)\right\rangle_{Z_1(Q)}\left\langle\phi^\alpha(\bx_1)\phi^\beta(\bx_2)\right\rangle_{Z_1(Q)}\Big].
\end{align}
Since $Z_1$ is Gaussian, Wick's theorem lets us write the four-point function in terms of two-point functions. One of the possible pairings cancels with the second term. Writing the remaining pairings in terms of $Q$ using the saddle point equations, we obtain
\begin{align}
\frac{\delta \left\langle\phi^\gamma(\bx_3)\phi^\delta(\bx_4)\right\rangle_{Z_1(Q)}}{\delta Q_{\alpha\beta}(\bx_1,\bx_2)}
&=\frac{1}{2}K_{\alpha\beta}(\bx_1,\bx_2)\Big[Q_{\gamma\alpha}(\bx_3,\bx_1)Q_{\delta\beta}(\bx_4,\bx_2) +Q_{\gamma\beta}(\bx_3,\bx_2)Q_{\delta\alpha}(\bx_4,\bx_1)\Big].
\end{align}
Finally, we obtain the expression for the Hessian after differentiating Eq.~\eqref{eq:actionFirstDeriv}:
\begin{align}
H_{\alpha\beta,\gamma\delta}(\bx_1,\bx_2, \bx_3,\bx_4)
&=\frac{1}{2}K_{\gamma\delta}(\bx_3,\bx_4)\Bigg\{\delta_{\gamma\alpha}\delta_{\delta\beta}\delta(\bx_3-\bx_1)\delta(\bx_4-\bx_2) \nonumber\\
&-\frac{1}{2}K_{\alpha\beta}(\bx_1,\bx_2)\Big[Q_{\gamma\alpha}(\bx_3,\bx_1)Q_{\delta\beta}(\bx_4,\bx_2)+Q_{\gamma\beta}(\bx_3,\bx_2)Q_{\delta\alpha}(\bx_4,\bx_1)\Big]\Bigg\}.
\end{align}
This form is general and valid for any saddle point. To test stability against RSB, we consider a translationally invariant variation
\begin{equation}
Q_{\alpha\beta}(\bx - \bx')
=
Q_{\alpha\beta}^{\mathrm{RS}}(\bx - \bx')
+\epsilon\,\eta_{\alpha\beta}(\bx - \bx'),
\end{equation}
where the RS form (Eq.~\ref{eq:QRSAnsatz}) is given in real space as
\begin{equation} Q^{\mathrm{RS}}_{\alpha \beta}(\bx - \bx') = C(\bx -\bx') \delta_{\alpha \beta} + q. \end{equation}
Here, $C(\bx -\bx') \equiv \langle \phi^{\alpha}(\bx) \phi^{\alpha}(\bx')\rangle_c$, obtained as the inverse Fourier transform of $S_\mathrm{conn}$. 
Since the first variation vanishes at the saddle, we have
\begin{align}
\mathcal{S}[Q^{\mathrm{RS}}+\epsilon\eta]
=& \mathcal{S}[Q^{\mathrm{RS}}]\nonumber + \epsilon^2 \Delta \mathcal S^{(2)} +\mathcal O(\epsilon^3). \\
\Delta \mathcal S^{(2)} \equiv& \frac{1}{2}
\sum_{\alpha\beta\gamma\delta}
\int_{\bx_1,\bx_2, \bx_3, \bx_4}
\eta_{\alpha\beta}(\bx_1 - \bx_2) 
H_{\alpha\beta,\gamma\delta}
(\bx_1,\bx_2,\bx_3,\bx_4)
\eta_{\gamma\delta}(\bx_3 - \bx_4)
\end{align}
As in Eq.~\eqref{eq:repliconeta}, we restrict to replicon perturbations 
\begin{equation} \label{eq:spatialReplicon} \eta_{\alpha\alpha}(\mathbf{r})=0,\qquad \eta_{\alpha\beta}(\mathbf{r})=\eta_{\beta\alpha}(-\mathbf{r}),\qquad \sum_{\beta}\eta_{\alpha\beta}(\mathbf{r})=0.\end{equation}
These satisfy $K_{\alpha\beta}(\mathbf{x},\mathbf{x}')
\eta_{\alpha\beta}(\mathbf{x}-\mathbf{x}')
=
\sigma^2\eta_{\alpha\beta}(\mathbf{x}-\mathbf{x}')$. We then have
\begin{align}
\Delta\mathcal S^{(2)}=&\frac{\sigma^2}{4}\sum_{\alpha\beta}\int_{\bx_1,\bx_2}\eta_{\alpha\beta}^2(\bx_1-\bx_2)\\
&-\frac{\sigma^4}{8}\sum_{\alpha\beta\gamma\delta}\int_{\bx_1,\bx_2,\bx_3,\bx_4}\!\!\!\eta_{\alpha\beta}(\bx_1-\bx_2)\left[Q^{\mathrm{RS}}_{\gamma\alpha}(\bx_3-\bx_1)Q^{\mathrm{RS}}_{\delta\beta}(\bx_4-\bx_2)+Q^{\mathrm{RS}}_{\gamma\beta}(\bx_3-\bx_2)Q^{\mathrm{RS}}_{\delta\alpha}(\bx_4-\bx_1)\right]\eta_{\gamma\delta}(\bx_3-\bx_4).
\end{align}
We evaluate the two terms in square brackets by contracting them with $\eta$ from right to left. For the first term, the contraction over $\delta$ gives
\begin{align}
\sum_{\delta}Q^{\mathrm{RS}}_{\delta\beta}(\bx_4-\bx_2)\eta_{\gamma\delta}(\bx_3-\bx_4)
&=C(\bx_4-\bx_2)\eta_{\gamma\beta}(\bx_3-\bx_4)+q\sum_{\delta}\eta_{\gamma\delta}(\bx_3-\bx_4)\nonumber\\
&=C(\bx_4-\bx_2)\eta_{\gamma\beta}(\bx_3-\bx_4),
\end{align}
where the second term vanishes by Eq.~\eqref{eq:spatialReplicon}. Repeating this procedure for the remaining contractions and reordering dummy integration variables, we find
\begin{align}
\Delta\mathcal S^{(2)}
=\frac{\sigma^2}{4}\sum_{\alpha\beta}\int_{\bx_1,\bx_2}\eta_{\alpha\beta}^2(\bx_1-\bx_2) -\frac{\sigma^4}{4}\sum_{\alpha\beta}\int_{\bx_1,\bx_2,\bx_3,\bx_4}\eta_{\alpha\beta}(\bx_1-\bx_2) C(\bx_3-\bx_1)C(\bx_4-\bx_2)\eta_{\alpha\beta}(\bx_3-\bx_4).
\end{align}
Using translational invariance to compute one of the space integrals in each term gains a factor of $V$. Fourier transforming then yields the final expression
\[
{
\Delta\mathcal S^{(2)}
=
\frac{\sigma^2 V}{4}
\sum_{\alpha\beta}
\int\frac{d^d\mathbf{k}}{(2\pi)^d}\,
\left[
1-\sigma^2S_{\mathrm{conn}}^2(k)
\right]
|\eta_{\alpha\beta}(\mathbf{k})|^2.
}
\]
The RS solution is thus stable if 
\begin{equation} \label{eq:sphericalStability}
    S_{\mathrm{conn}}(k) \leq \frac{1}{\sigma}.
\end{equation}
for all $k$. In the high temperature phase, where $S_\mathrm{conn} = S_{\mathrm{HT}}$, the inequality is strict: Equation~\eqref{eq:HTsphericalSF} satisfies Eq.~\eqref{eq:sphericalStability} because $r+up > 2\sigma$  for $r>r_c$. At the critical point and throughout the glass phase, $S_\mathrm{conn}(k)$ assumes its critical form $S_c$, and $S_\mathrm{c}(0) = 1/\sigma$. The uniform replicon mode is thus marginal in the glass phase, but there is never an AT instability.

\subsection{Relation to the spherical spin glass in $d=0$}
In $d=0$, the model reads
\begin{equation} \label{eq:zerodSpherical}
\beta\mathcal H
=
\frac{r}{2}\,\vec{\phi}^{\,2}
+\frac{1}{2}\,\vec{\phi}^{\,T}J\vec{\phi}
+\frac{u}{4N}\left(\vec{\phi}^{\,2}\right)^2.
\end{equation}
According to the previous analysis, as
$N\to\infty$, the physical overlap $\widetilde Q_{\alpha\beta}
=\frac{1}{N}\vec\phi^{\,\alpha}\cdot\vec\phi^{\,\beta}$ concentrates on its saddle-point value $Q_{\alpha \beta}$, with
fluctuations that vanish at large $N$. In particular,
\[
\frac{1}{N}\vec\phi^{\,2}
\rightarrow Q_{\alpha\alpha}
\equiv p.
\]
The field is thus restricted at large $N$ to a sphere
$\vec\phi^{\,2}=Np$. Writing 
\[\vec{\phi} = \sqrt{p} \vec{s}, \qquad (\vec{s})^2 = N,\]
gives
\begin{equation} \label{eq:sphericalSpinGlass}
\beta\mathcal H
=
N\left(\frac{r}{2}p+\frac{u}{4}p^2\right)
+\frac{p}{2}\vec s^{\,T}J\vec s.
\end{equation}
For fixed \(p\), the first term is constant, while the remaining angular dependence is that of the spherical spin glass~\cite{kosterlitz1976spherical}. In the original model, $p$ is determined self-consistently~\footnote{This $p$ is not to be confused with the $p$-spin spherical spin glass; in that language, Eq.~\eqref{eq:sphericalSpinGlass} is a 2-spin spherical spin glass.}
\bibliography{bibliography}